# An Ecologically-Informed Deep Learning Framework for Interpretable and Validatable Habitat Mapping

Iván Felipe Benavides-Martínez[1,2,*], Cristiam Victoriano Portilla-Cabrera[3], Katherine E. Mills[1], Claire Enterline[1], José Garcés-Vargas[4,5], Andrew J. Allyn[1,6], Auroop R Ganguly[2,7]

[1]Gulf of Maine Research Institute, Portland, Maine, USA

[2]Artificial Intelligence for Climate and Sustainability, The Institute for Experiential Artificial Intelligence, Northeastern University, Portland, ME, USA

[3]Grupo de Investigación en Recursos Hidrobiológicos, Departamento de Ingeniería, Universidad Nacional de Colombia– Sede Palmira, Carrera 32 No. 12-00, Palmira, Valle del Cauca, Colombia

[4]Instituto de Ciencias Marinas y Limnológicas, Universidad Austral de Chile, 5090000, Valdivia, Chile

[5]Centro FONDAP de Investigación en Dinámica de Ecosistemas Marinos de Altas Latitudes (IDEAL), 5090000, Valdivia, Chile

[6]Intercampus Marine Science Program, University of Massachusetts Amherst, Amherst, Massachusetts, USA

[7]Sustainability and Data Sciences Laboratory, Northeastern University, Boston, MA, USA

**Abstract**

Benthic habitat is challenging due to the environmental complexity of the seafloor, technological limitations, and elevated operational costs, especially in under-explored regions. This generates knowledge gaps for the sustainable management of hydrobiological resources and their nexus with society. We developed ECOSAIC (Ecological Compression via Orthogonal Specialized Autoencoders for Interpretable Classification), an Artificial Intelligence framework for automatic classification of benthic habitats through interpretable latent representations using a customizable autoencoder. ECOSAIC compresses n-dimensional feature space by optimizing specialization and orthogonality between domain-informed features. We employed two domain-informed categories: biogeochemical and hydrogeomorphological, that together integrate biological, physicochemical, hydrological and geomorphological, features, whose constraints on habitats have been recognized in ecology for a century. We applied the model to the Colombian Pacific Ocean and the results revealed 16 benthic habitats, expanding from mangroves to deep rocky areas up to 1000 m depth. The candidate habitats exhibited a strong correspondence between their environmental constraints, represented in latent space, and their expected species composition. This correspondence reflected meaningful ecological associations rather than purely statistical correlations, where the habitat's environmental offerings align semantically with the species' requirements. This approach could improve the management and conservation of benthic habitats, facilitating the development of functional maps that support marine planning, biodiversity conservation and fish stock assessment. We also hope it provides new insights into how ecological principles can inform AI frameworks, particularly given the substantial data limitations that characterize ecological research.

---

*Corresponding author.
Email addresses: Iván Felipe Benavides-Martínez (fbenavides@gmri.org), Cristiam Victoriano Portilla-Cabrera (cvportillac@unal.edu.co), Katherine E. Mills (kmills@gmri.org), Claire Enterline (centerline@gmri.org), José Garcés-Vargas (jose.garces.vargas@gmail.com), Andrew J. Allyn (aallyn@gmri.org), Auroop R Ganguly (a.ganguly@northeastern.edu)

**Keywords:** Autoencoder, semantic compression, deep learning, benthic habitats, Colombian Pacific

## 1. Introduction

Benthic marine ecosystems constitute the single largest ecosystem on Earth and provide ecosystem services of ecological, economic, and social importance (Hilmi et al., 2023; Tsikopoulou et al., 2024). These marine benthic habitats represent physically distinct areas of the seafloor inhabited by particular organisms, serving as vital links between shallow productive environments and the abyssal bottom (Yang et al., 2025). Together, marine ecosystems and their associated biodiversity sustain life on Earth including the maintenance of global oxygen and carbon cycles, the production of food and energy, and the support of human well-being (Duarte et al., 2022). Given that empirical studies have demonstrated the role of species diversity in sustaining ecosystem processes, particularly in these marine benthic ecosystems (Covich et al., 2004), it is evident that benthic habitat mapping is increasingly recognized as an essential tool for informing ecosystem-based marine management, representing an indispensable tool for marine conservation and spatial planning (Galparsoro et al., 2014).

Although benthic habitat mapping is a fundamental tool in marine resource management, considerable technical and logistical challenges still exist that limit its implementation at large scale. Fundamentally, accurate mapping presents substantial challenges due to the vastness and complexity of marine environments, compounded by technological limitations and high resource demands (Yang et al., 2025). On one hand, the application of remote sensing techniques is restricted to shallow coastal waters due to limited light penetration, leaving most of the seafloor beyond the reach of these techniques (Brown et al., 2011). On the other hand, limitations of multibeam sonar systems include navigational hazards, inability to collect data in shallow waters less than 15 meters, and presence of noise and interference that limit data quality (Costa et al., 2009). Additionally, autonomous underwater vehicles still face significant challenges including battery limitations, submarine communications, and navigation (Sun et al., 2021). As a result of these technical limitations, traditional methods for marine cartography require the use of specialized vessels and advanced technological equipment, which implies high operational costs associated with maintenance, fuel, qualified personnel, and insurance, limiting the frequency and scope of maritime explorations (National Research Council, 2009). This problem is clearly evident in countries with large exclusive economic zones such as Spain, Portugal, and France that have less mapped areas due to the major technical and economic challenges involved in charting bathyal and abyssal zones (Galparsoro et al., 2014). In the regional context of Colombia, especially in the Colombian Pacific, these challenges are further aggravated by institutional limitations and the geographic complexity of the region, which hinder research and the development of maritime knowledge in deep areas (RAP Pacífico, 2022).

In response to these identified limitations, the scientific community has developed advanced machine learning techniques to optimize seafloor mapping. Specifically, supervised learning methods applied to terrestrial remote sensing provide a promising approach for marine habitat characterization (Hasan et al., 2012). Among the most successful algorithms, Random Forest, Classification Tree Analysis, and Support Vector Machine stand out, which have been successfully integrated with in situ benthic data (Wicaksono et al., 2019). In parallel, Convolutional Neural Networks and Transformer-based models provide transfer learning frameworks to classify deep-water habitats (Game et al., 2024). Likewise, automatic and semi-automatic classification approaches have demonstrated effectiveness in high-resolution three-dimensional mapping using Structure from Motion algorithms (Mohamed et al., 2020). More recently, classification techniques have been developed to leverage high-resolution Synthetic Aperture Sonar (SAS) images for seafloor mapping (Sørensen et al., 2025). In general terms, machine learning algorithms improve accuracy and efficiency by processing large datasets and extracting complex patterns (Evans et al., 2025), while marine remote sensing techniques provide necessary tools to map and monitor ecosystems more effectively (Costa et al., 2009).

Despite the efforts made and the technological advances described, fundamental limitations still persist that compromise the practical applicability of these emerging techniques in marine management contexts. First, current models for seafloor mapping lack designs oriented toward optimizing the biological interpretability of results, assuming uniform species sensitivity to abiotic variables (Yang et al., 2025). Additionally, objective labels do not exist for many seafloor habitats, biological communities, and substrate types, representing a major difficulty for the development of Deep Learning models (Marburg & Bigham, 2016). From a methodological perspective, the lack of standard calibration of sonar systems produces backscatter measurements relative to each survey, presenting challenges for mapping in areas with multiple surveys (Misiuk et al., 2020). Furthermore, sonar images present noise and interference, and the high cost of field sampling results in limited samples that restrict classification accuracy (Wang et al., 2025). Among the specific technical challenges is the mitigation of acoustic shadowing in high-resolution image classification techniques (Sørensen et al., 2025). Finally, the reliability of results depends on the quality of habitat maps and expert judgment evaluation, which can be biased (Galparsoro et al., 2014), while the absence of semantically guided compression represents a critical gap, making it difficult to validate whether identified spatial units are ecologically significant.

To address the identified limitations and advance toward benthic cartography with robust ecological interpretability, this study proposes an integrated workflow structured in five sequential phases: (i) delimitation and characterization of the study area on the seafloor of the Colombian Pacific Ocean, spanning the bathymetric range of 0 to 1,000 meters depth; (ii) compilation and preparation of multiple environmental variables representative of the benthic environment, including physical, chemical, and geomorphological characteristics relevant to habitat structuring; (iii) implementation of a modular autoencoder architecture specifically designed for semantic compression of heterogeneous environmental variables through independent parallel processing, employing specialized encoders for each functional category that generate orthogonal latent representations, minimizing redundancy between domains while preserving ecological significance during dimensional reduction and facilitating the identification of domain-specific functional patterns without compromising the system's integral reconstruction capacity; (iv) unsupervised segmentation through self-organizing maps (Self-Organizing Maps, SOM) on the reduced latent space, grouping benthic habitat units into a topologically preserved grid structure that reduces the inherent subjectivity of manual classifications and improves spatial consistency; and (v) validation through two complementary approaches: structural validation of compression, relating latent variables to original environmental variables through scatter plots stratified by habitat class to corroborate congruence with natural environmental gradients, and ecological validation, through spatial intersection of habitat units with georeferenced species occurrence records from validated biological databases, verifying the correspondence between modeled environmental units and observed distribution patterns of benthic biota. This integrated methodology improves the ecological interpretability of benthic habitats by projecting them onto a reduced but functionally relevant feature space, facilitating the generation of cartography with explicit biological significance and potential applicability in ecosystem management.

## 2. Materials and Methods

The methodological process was structured systematically in five sequential stages designed to identify and spatially validate the benthic habitats associated with the Colombian Pacific Ocean. Initially, the study area was delimited and described considering physical, geological, hydrological, and geomorphological criteria (2.1). Subsequently, environmental variables from the seafloor were compiled and processed, including biogeochemical and hydrogeomorphological parameters (2.2). In the third stage, an autoencoder architecture specifically designed for dimensional reduction and grouping of environmental variables was implemented, enabling the identification of latent patterns in multidimensional space (2.3). From the model results, a raster classification map of habitats was generated that spatially integrated the identified groups, providing a continuous cartographic representation of benthic units (2.4). Finally, the habitat map was validated by relating latent variables to the 33 original environmental variables through scatter plots labeled according to habitat class, corroborating spatial coherence with oceanographic patterns of the Colombian Pacific. Additionally, ecological validation was performed through spatial intersection of habitats with known species occurrence records, allowing verification of real ecological correspondence between habitat units and observed species distribution patterns (2.5).

### 2.1 Study Area

The study area comprised the seafloor associated with the Colombian Pacific Ocean, encompassing approximately 19,008 km² between zero and 1,000 meters depth (Figure 1). This zone constitutes a distinctive geomorphological unit within the extreme eastern portion of the Eastern Tropical Pacific, characterized by high primary productivity, pronounced bathymetric gradients, and high sediment inputs transported primarily by the Patía, Mira, Baudó, and Atrato rivers (Restrepo & Kjerfve, 2004). This region is characterized by a bipartite coastal morphology reflecting its complex tectonic evolution: to the north, from the border with Panama to Cape Corrientes, elevated coasts with cliffs composed of Tertiary rocks of the Baudó Sierra predominate, with adjacent mountain formations reaching up to 100 m in height near the coast; while to the south, from Cape Corrientes to the border with Ecuador, a coast with predominantly flat relief is present, constituted by extensive deltaic plains and wide mangrove areas aligned parallel to the coastline (Oviedo-Barrero et al., 2020).

The current geomorphological configuration of the Colombian Pacific is the result of intensive Quaternary sedimentary processes fed by more than 200 short rivers characterized by high discharge and significant sediment transport, with total discharge exceeding 9,000 m³/s and a sedimentary input of 96 million tons annually (Díaz-Merlano, 2007; Restrepo et al., 2002). Predominant sediments in this coastal zone are gravels, sands, and silts of fluvial origin, deposited during the Quaternary, forming a coastal plain that shows localized interruptions by small cliffs in the bays of Málaga, Buenaventura, and Tumaco, associated with a series of anticlines—geological structures formed by the folding of rock layers in the Earth's crust—formed by the tectonic activity of the Andes toward the end of the Early Pleistocene (Oviedo-Barrero et al., 2020). This geodynamics continues to influence the morphological evolution of the region, where the most important rivers such as Baudó, San Juan, Naya, Guapi, Iscuandé, Patía, Mataje, Micay, and Mira form individual or multiple deltas that significantly contribute to the shaping of the coastal landscape (Díaz-Merlano, 2007).

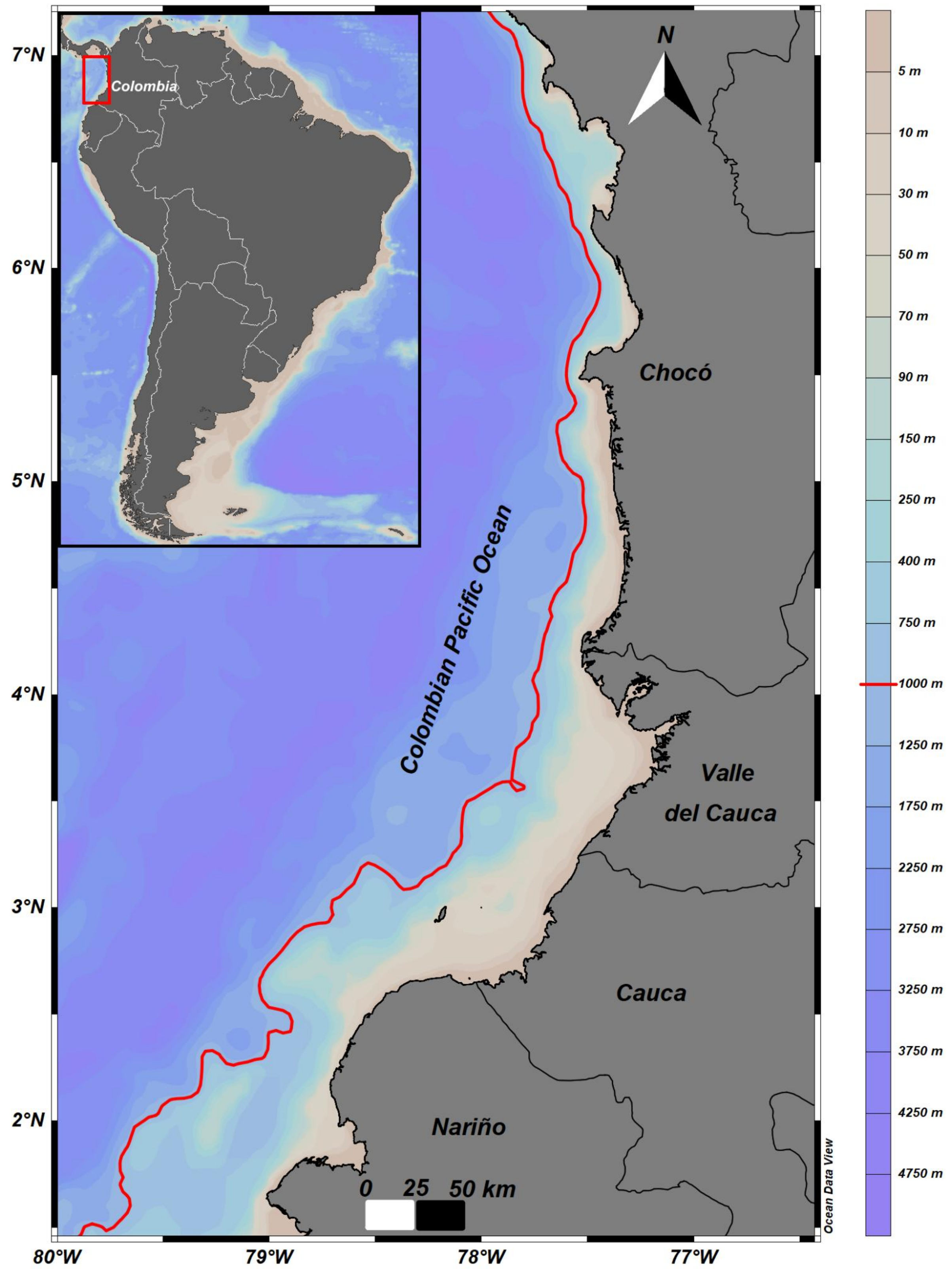

Figure 1. Delimitation of the study area (red line) up to the 1,000 m isobath in the Colombian Pacific Ocean. The upper left inset shows the location of the area in the South American context.

## 2.2 Compilation of Seafloor Environmental Variables

### 2.2.1 Acquisition of Variables from Official Sources

The environmental database was constructed by integrating 33 seafloor variables (15 biogeochemical and 18 hydrogeomorphological) obtained from open-access official sources, including Bio-ORACLE (Assis et al., 2024), Global Marine Environmental Dataset–GMED (Basher et al., 2018), Copernicus Climate Change Service (Hersbach et al., 2023), derivatives from bathymetry using the TASSE Toolbox package (Lecours et al., 2016, 2017), physicochemical layers predicted by Rosales-Estrella et al., (2025), as well as layers adapted from the Instituto Geográfico Agustín Codazzi - IGAC, (2016) (Table 1).

#### 2.2.2 Feature Engineering

Since environmental variables were acquired at different spatial resolutions, the resolution was standardized to 0.04° through bilinear interpolation to ensure spatial consistency across layers, following established protocols for processing oceanographic data from multiple sources (Benavides Martínez et al., 2024). Given the presence of missing values in some variables, particularly in extensive coastal strips of the study area, a spatial inpainting approach based on Conditional Generative Adversarial Networks (C-GAN) was implemented. This method is appropriate for inpainting geospatial data because it learns complex and nonlinear patterns, maintains spatial coherence between original and reconstructed areas, and efficiently handles missing data with irregular shapes, guaranteeing realistic and spatially consistent reconstructions (Li & Cao, 2025; Zhao et al., 2021). C-GANs have demonstrated specific superiority over traditional interpolation methods in climate and oceanographic data applications, providing more accurate reconstructions and better preserving the natural variability of geospatial time series (Ham et al., 2024).

The C-GAN architecture employed integrated a U-Net type generator with residual connections and a convolutional discriminator, conditioning the generator through auxiliary channels that included a study area mask, a Euclidean distance map to edges of known data, and contextual geomorphological variables derived from bathymetry (Table 1). Adversarial training optimized a combined loss function integrating L1 reconstruction loss (weighting factor: 100) and binary adversarial loss, using Adam optimizers with differentiated learning rates (generator: 0.0001; discriminator: 0.00005). The inpainting results obtained through the C-GAN model were comparatively evaluated against U-Net and Kriging methods using a longitudinal cross-validation scheme with three random partitions (80% of the longitudinal strips for training and 20% remaining for testing). Performance quantification was based on four complementary metrics that evaluated different aspects of reconstruction quality: Mean Absolute Error (L1), which measures the average magnitude of prediction errors providing robustness to outliers; Mean Squared Error (L2), which disproportionately penalizes large errors and is sensitive to model variance (Hodson, 2022); Peak Signal-to-Noise Ratio (PSNR), which quantifies reconstruction fidelity in terms of the ratio between the maximum possible signal power and the noise affecting its representation; and Structural Similarity Index (SSIM), which evaluates the preservation of local spatial patterns considering changes in luminance, contrast, and structure (Horé & Ziou, 2010). This multidimensional metric ensemble ensured comprehensive evaluation of each method's capacity to preserve both numerical accuracy and spatial coherence of reconstructed environmental variables (Supplementary Material: Table S1).

Upon completion of spatial alignment and inpainting processes, systematic value extraction by centroid was performed for all environmental variables, generating a structured data matrix. Subsequently, tabulated environmental variables were scaled using RobustScaler to mitigate the influence of outliers, followed by min-max normalization to the standard range to ensure comparability across variables, following established best practices for preprocessing multivariate oceanographic data (Heimbach et al., 2025). Afterward, specific transformations adapted to the ecological characteristics of each variable were implemented: sigmoid transformations for pH,

Michaelis-Menten kinetics for inorganic nutrients (phosphates, iron, nitrate, and silicate), and exponential attenuation functions for light variables (bottom light and photosynthetically active radiation). Biological variables, including chlorophyll-a, phytoplankton, oxygen utilization, and primary productivity, were subjected to negative exponential transformations to capture their nonlinear dynamics. Additionally, double sigmoid transformations were applied to dissolved oxygen and salinity, recognizing their complex spatial gradients and species-specific ecological thresholds, thus differentiating from more generalist transformation approaches. On the other hand, differentiated transformations were implemented for hydrogeomorphological variables: cosine functions for bathymetric aspect, current velocity, and wind; hyperbolic tangent function for east, north orientations and longitudinal river influence; fractal power functions for runoff, distance to land, and local mean terrain; and arctangent functions for relative position, slope, and standard deviation of terrain. Substrate hardness and particle size were transformed through rational functions to model their characteristic nonlinear distributions in marine sediments. The IRBM transformation implemented a hybrid mixed function combining sigmoidal and arctangent components (Table 1).

Finally, controlled Gaussian noise with differentiated standard deviations by category was incorporated: 0.0008 for biogeochemical variables and 0.002 for hydrogeomorphological variables, reflecting both the greater sensitivity of biogeochemical variables and the different scales of natural variability inherent to each category. This addition of controlled noise fulfills three essential methodological purposes based on established principles of regularization in machine learning (Baig et al., 2023): (i) prevention of overfitting through the introduction of realistic variability that avoids exact memorization of training patterns, (ii) improvement of model generalization by simulating natural uncertainties present in real oceanographic data, and (iii) implicit regularization that forces latent representations to capture robust patterns rather than spurious features.

Table 1. Seafloor environmental variables included in the study. This table includes variable names, applied transformations, and data sources. Variables marked with (*) represent biogeochemical parameters; remaining variables represent hydrogeomorphological parameters.

| Variable | Applied Transformation | Source |
|---|---|---|
| *Phosphate [mol·m$^{-3}$] | Michaelis-Menten $f(X) = \frac{X^{2.80}}{0.090+X^{2.80}}$ | Bio-ORACLE |
| *Iron [µmol·m$^{-3}$] | Michaelis-Menten $f(X) = \frac{X^{4.00}}{0.170+X^{4.00}}$ | Bio-ORACLE |
| *Bottom light [Einstein·m$^{-2}$·day$^{-1}$] | Exponential attenuation $f(X) = e^{-3.40(1-X)}$ | Bio-ORACLE |
| *Nitrate [mol·m$^{-3}$] | Michaelis-Menten $f(X) = \frac{X^{2.80}}{0.090+X^{2.80}}$ | Bio-ORACLE |
| *Dissolved oxygen [mmol·m$^{-3}$] | Double sigmoid | Bio-ORACLE |
| | $f(X) = \frac{0.7}{1+e^{-12.0(X-0.3)}} + \frac{0.3}{1+e^{-7.0(X-0.7)}}$ | |
| *pH | Sigmoid<br>$f(X) = \frac{1}{1+e^{-16.0(X-0.47)}}$ | GMED |
| *Salinity [PSU] | Double sigmoid<br>$f(X) = \frac{0.7}{1+e^{-15.0(X-0.3)}} + \frac{0.3}{1+e^{-8.5(X-0.7)}}$ | Bio-ORACLE |
| *Silicate [mol·m$^{-3}$] | Michaelis-Menten<br>$f(X) = \frac{X^{2.80}}{0.090+X^{2.80}}$ | Bio-ORACLE |
| *Temperature [°C] | Double sigmoid<br>$f(X) = \frac{0.7}{1+e^{-6.0(X-0.3)}} + \frac{0.3}{1+e^{-4.0(X-0.7)}}$ | Bio-ORACLE |
| *Photosynthetically active radiation [Einstein·m$^{-2}$·day$^{-1}$] | Exponential attenuation<br>$f(X) = e^{-4.60(1-X)}$ | Bio-ORACLE |
| *Organic matter [%] | Sigmoid + quadratic<br>$f(X) = 0.6 \times \frac{1}{1+e^{-11.0(X-0.4)}} + 0.4 \times X^2$ | Rosales-Estrella et al. (2025) |
| *Chlorophyll-a [mg·m$^{-3}$] | Negative exponential<br>$f(X) = 1 - e^{-8.20X^{1.90}}$ | Bio-ORACLE |
| *Phytoplankton [µmol·m$^{-3}$] | Negative exponential<br>$f(X) = 1 - e^{-5.80X^{1.50}}$ | Bio-ORACLE |
| *Bottom utilized oxygen [ml·l$^{-1}$] | Double sigmoid<br>$f(X) = \frac{0.7}{1+e^{-6.0(X-0.3)}} + \frac{0.3}{1+e^{-4.0(X-0.7)}}$ | GMED |
| *Primary productivity [g·m$^{-3}$·day$^{-1}$] | Negative exponential<br>$f(X) = 1 - e^{-10.00X^{2.20}}$ | Bio-ORACLE |
| Aspect [degrees] | Cosine<br>$f(X) = 0.5(1 + \cos(2.600\pi X + 6.00))$ | TASSE Toolbox |
| Eastness | Hyperbolic tangent $f(X) = \tanh(3.00(X - 0.5)) \times 0.5 + 0.5$ | TASSE Toolbox |
| Longitudinal river influence | Hyperbolic tangent $f(X) = \tanh(3.80(X - 0.5)) \times 0.5 + 0.5$ | Adapted from IGAC (2016) |
| Benthic current velocity [m·s⁻¹] | Cosine $f(X) = 0.5(1 + \cos(2.600\pi X + 6.00))$ | Bio-ORACLE |
| Runoff [m] | Fractal power $f(X) = (1 - X)^{0.60}$ | Copernicus Climate Change Service |
| 10 m v-component of wind [m·s⁻¹] | Cosine $f(X) = 0.5(1 + \cos(1.550\pi X + 2.50))$ | Copernicus Climate Change Service |
| Land distance [degrees] | Arctangent $f(X) = \frac{2}{\pi}\arctan(4.10X)$ | GMED |
| Local mean terrain [m] | Fractal power $f(X) = (1 - X)^{1.20}$ | TASSE Toolbox |
| Northness | Hyperbolic tangent $f(X) = \tanh(3.80(X - 0.5)) \times 0.5 + 0.5$ | TASSE Toolbox |
| Terrain Position Index | Arctangent $f(X) = \frac{2}{\pi}\arctan(3.50X)$ | TASSE Toolbox |
| Slope [degrees] | Arctangent $f(X) = \frac{2}{\pi}\arctan(5.60X)$ | TASSE Toolbox |
| Terrain roughness (elevation std. dev.) [m] | Logarithmic $f(X) = \frac{\log(1+9X)}{\log(10)}$ | TASSE Toolbox |
| Terrain surface | Logarithmic $f(X) = \frac{\log(1+9X)}{\log(10)}$ | TASSE Toolbox |
| Substrate hardness [%] | Rational mixed $f(X) = 0.700 + (1 - 0.700) \times \frac{X^{3.70}}{0.3+X^{3.70}}$ | Rosales-Estrella et al. (2025) |
| Particle size [phi] | Rational mixed $f(X) = 0.750 + (1 - 0.750) \times \frac{X^{3.90}}{0.3+X^{3.90}}$ | Rosales-Estrella et al. (2025) |
| Euphotic layer depth [m] | Exponential attenuation $f(X) = e^{-3.80(1-X)}$ | GMED |
| Light attenuation coefficient [$m^{-1}$] | Exponential attenuation $f(X) = e^{-3.40(1-X)}$ | GMED |
| Radial Influence of Mangrove Biomass (IBMR) | Hybrid mixed $f(X) = 0.6 \times \frac{1}{1+e^{-4.00(X-0.4)}} + 0.4 \times \frac{2}{\pi} \arctan(2.00X)$ | Adapted from Selvaraj & Gallego Pérez (2023) |

### 2.3 Autoencoder Architecture

We developed an autoencoder architecture termed ECOSAIC (Ecologically-informed Compression through Orthogonal Specialized Autoencoders for Interpretable Classification), specifically designed for semantic compression of multiple environmental variable categories through domain specialization and orthogonality. Multi-objective optimization approaches in machine learning applied to marine systems have demonstrated superior capacity to balance multiple conflicting criteria while preserving critical information during complex dimensional reduction processes (Bolton & Zanna, 2019). This architecture enabled capture of distinctive characteristics of each functional category while minimizing inter-domain redundancies, addressing limitations identified in traditional dimensional reduction methods applied to oceanographic data (Falasca et al., 2024).

Initially, we considered compressing three functional categories: hydrodynamic, physicochemical, and biological. However, given the high cross-correlation observed between physicochemical and biological variables, we reconfigured the architecture to process two integrated categories: (i) biogeochemical, integrating biological and physicochemical variables (n=15), generating the latent representation LE1; and (ii) hydrogeomorphological, combining hydrodynamic and geomorphological variables (n=18), producing the latent representation LE2 (Figure 2).

The architecture implemented specialized encoders with different depths and capacities optimized for each functional domain, following principles of modular autoencoder architectures that have demonstrated effectiveness in compressing complex system dynamics (Baig et al., 2023) (Figure 2). The biogeochemical encoder (LE1) utilized a deep 5-layer dense architecture with progressive descending neuronal configuration (224-160-96-64-32 neurons), culminating in a one-dimensional latent representation, with variable dropout rates (0.05–0.30) and specialized activation functions (ReLU, SiLU, ELU, GELU, SELU) to capture the complexity of biogeochemical interactions among physical, chemical, and biological variables. The hydrogeomorphological encoder (LE2) employed a 4-layer architecture with expanded initial neuronal configuration (224-112-112-64 neurons), equally generating a one-dimensional latent representation, with more aggressive dropout (0.12–0.22) for robust regularization and activations (SELU, ELU, ReLU, SiLU) designed to process geomorphological and hydrodynamic features. The latent layers incorporated L1-L2 regularization (LE1: $\lambda_{L1}$=0.08, $\lambda_{L2}$=0.22; LE2: $\lambda_{L1}$=0.10, $\lambda_{L2}$=0.13) with directional RandomNormal kernel initialization toward positive values ($\mu$=0.4, $\sigma$=0.1) for LE1 and negative values ($\mu$=−0.6, $\sigma$=0.12) for LE2, facilitating spatial separation in the latent space and promoting inter-domain orthogonality. The decoders implemented mirror architectures with gradual expansion (LE1: 64-96-112-176-256 neurons; LE2: 64-72-104-208 neurons) from one-dimensional latent representations to original dimensions (15 and 18 variables respectively), utilizing sigmoid activation in the final layer to maintain consistency with the [0,1] range of normalized data. The architecture applied batch normalization after each dense layer with momentum 0.99 and $\varepsilon$ = 0.001 to stabilize training and accelerate convergence, following established best practices for deep autoencoder architectures (Üstek et al., 2024).

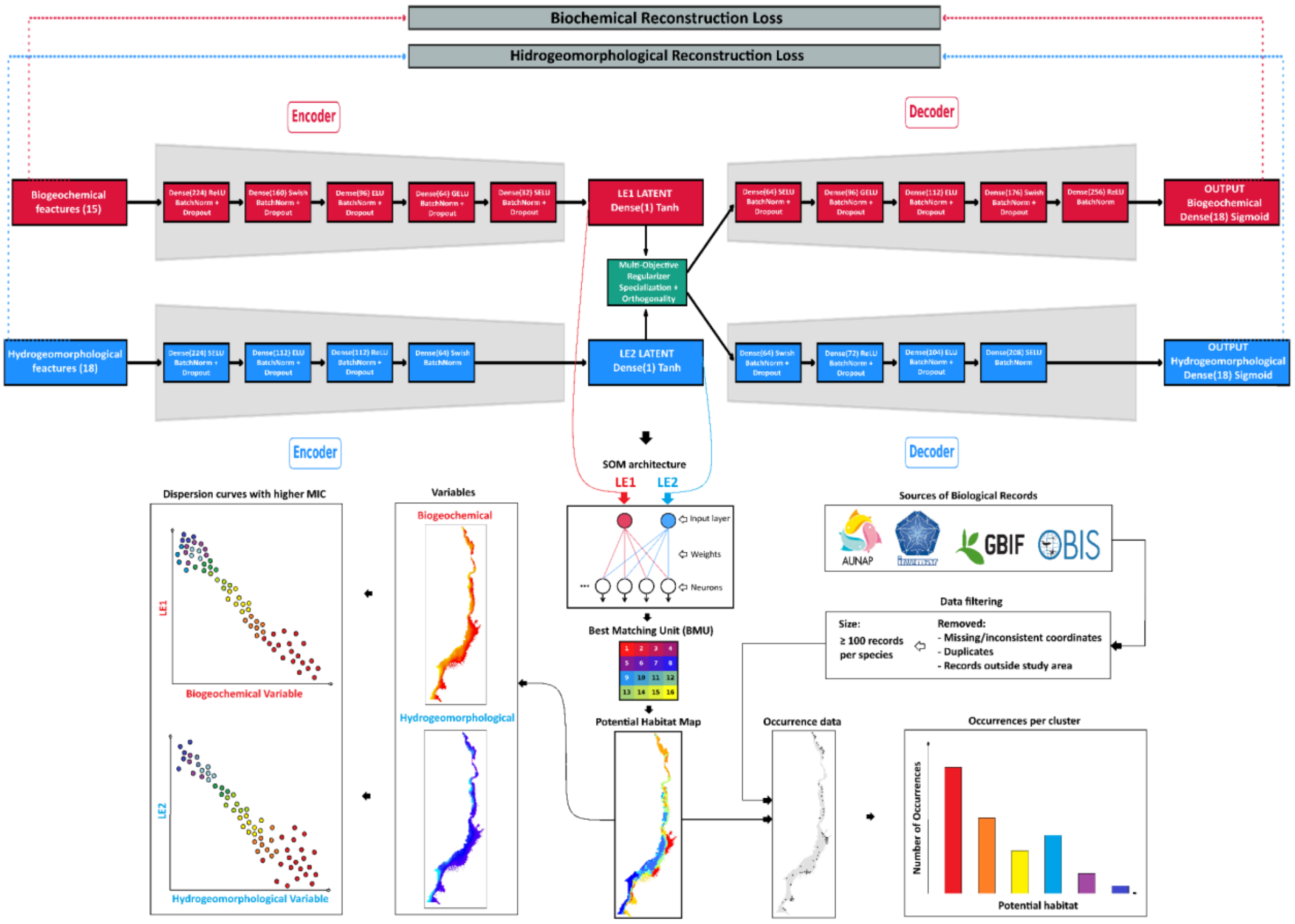

Figure 2. Overview of the ECOSAIC and Self-Organizing Map methodology for benthic habitat classification. The workflow integrates semantic compression of oceanographic variables through specialized encoders, generation of orthogonal latent embeddings, and ecological validation using species occurrence data.

### 2.3.1 Mathematical Formulation of the ECOSAIC Model

The ECOSAIC architecture processes environmental variables through domain-specific pathways. Table 2 defines the mathematical notation used throughout this section.

Table 2. Mathematical notation and variable definitions.This table defines all mathematical symbols used in the ECOSAIC architecture formulation.

| Symbol | Description | Dimension |
|---|---|---|
| $\mathbf{x}_{\text{biogeo}}$ | Biogeochemical input variables | $\mathbb{R}^{15}$ |
| $\mathbf{x}_{\text{hydro}}$ | Hydrogeomorphological input variables | $\mathbb{R}^{18}$ |
| $\mathbf{h}_i^{\text{biogeo}}$ | Hidden layer $i$ of biogeochemical encoder | Varies |
| $\mathbf{h}_i^{\text{hydro}}$ | Hidden layer $i$ of hydrogeomorphological encoder | Varies |
| $\mathbf{z}_{\text{biogeo}}$ | Biogeochemical latent representation | $\mathbb{R}^{1}$ |
| $\mathbf{z}_{\text{hydro}}$ | Hydrogeomorphological latent representation | $\mathbb{R}^{1}$ |
| $\mathbf{z}$ | Combined orthogonal latent space | $\mathbb{R}^{2}$ |
| $\mathbf{d}_i^{\text{biogeo}}$ | Decoder layer $i$ for biogeochemical branch | Varies |
| $\mathbf{d}_i^{\text{hydro}}$ | Decoder layer $i$ for hydrogeomorphological branch | Varies |
| $\hat{\mathbf{x}}_{\text{biogeo}}$ | Reconstructed biogeochemical variables | $\mathbb{R}^{15}$ |
| $\hat{\mathbf{x}}_{\text{hydro}}$ | Reconstructed hydrogeochemical variables | $\mathbb{R}^{18}$ |
| $\mathbf{W}_i$, $\mathbf{b}_i$ | Weight matrices and bias vectors for layer $i$ | Varies |
| $\text{BN}(\cdot)$ | Batch normalization (momentum $= 0.99, \varepsilon = 0.001$) | - |
| $\mathcal{L}$ | Loss function components | - |

The ECOSAIC architecture independently processes biogeochemical ($\mathbf{x}_{\text{biogeo}} \in \mathbb{R}^{15}$) and hydrogeomorphological ($\mathbf{x}_{\text{hydro}} \in \mathbb{R}^{18}$) variables through specialized encoders that converge into a two-dimensional orthogonal latent space $\mathbf{z} \in \mathbb{R}^2$, followed by independent decoders for data reconstruction (Figure 2). This dual-branch design enables domain-specific feature extraction while enforcing functional independence through orthogonality constraints, addressing the distinct temporal and spatial dynamics of biogeochemical versus hydrogeomorphological processes in marine systems.

Biogeochemical encoder: the biogeochemical encoder transforms 15 input features through five sequential dense layers with progressively decreasing dimensionality, capturing hierarchical patterns from physicochemical and biological interactions. Each transformation applies batch normalization (BN) before activation to stabilize training, followed by dropout regularization to prevent overfitting.

The first layer expands the input to a high-dimensional representation:

$$\mathbf{h}_1^{\text{biogeo}} = \text{ReLU}(\text{BN}(\mathbf{W}1^{\text{biogeo}}\mathbf{x}\text{biogeo} + \mathbf{b}_1^{\text{biogeo}})), \quad \mathbb{R}^{15} \to \mathbb{R}^{224} \tag{1}$$

Subsequent layers progressively compress the representation while extracting increasingly abstract features through diverse activation functions optimized for capturing non-linear biogeochemical relationships:

$$\mathbf{h}_2^{\text{biogeo}} = \text{SiLU}(\text{BN}(\mathbf{W}_2^{\text{biogeo}}\mathbf{h}_1^{\text{biogeo}} + \mathbf{b}_2^{\text{biogeo}})), \quad \mathbb{R}^{224} \to \mathbb{R}^{160} \tag{2}$$

$$\mathbf{h}_3^{\text{biogeo}} = \text{ELU}(\text{BN}(\mathbf{W}_3^{\text{biogeo}}\mathbf{h}_2^{\text{biogeo}} + \mathbf{b}_3^{\text{biogeo}})), \quad \mathbb{R}^{160} \to \mathbb{R}^{96} \tag{3}$$

$$\mathbf{h}_4^{\text{biogeo}} = \text{GELU}(\text{BN}(\mathbf{W}_4^{\text{biogeo}}\mathbf{h}_3^{\text{biogeo}} + \mathbf{b}_4^{\text{biogeo}})), \quad \mathbb{R}^{96} \to \mathbb{R}^{64} \tag{4}$$

$$\mathbf{h}_5^{\text{biogeo}} = \text{SELU}(\text{BN}(\mathbf{W}_5^{\text{biogeo}}\mathbf{h}_4^{\text{biogeo}} + \mathbf{b}_5^{\text{biogeo}})), \quad \mathbb{R}^{64} \to \mathbb{R}^{32} \tag{5}$$

The encoder culminates in a one-dimensional latent representation using hyperbolic tangent activation to bound the feature space to [-1, 1], with L1 and L2 regularization applied to promote sparsity and prevent overfitting:

$$\mathbf{z}\text{biogeo} = \tanh(\mathbf{W}\text{LE1}\mathbf{h}5^{\text{biogeo}} + \mathbf{b}\text{LE1}), \quad \mathbb{R}^{32} \to \mathbb{R}^{1} \tag{6}$$

where BN(·) denotes batch normalization with momentum 0.99 and $\varepsilon = 0.001$. Dropout layers with rates ranging from 0.05 to 0.30 are applied after each batch normalization layer to prevent overfitting, with higher rates in deeper layers. The latent layer incorporates L1-L2 regularization ($\lambda_{L1} = 0.08$, $\lambda_{L2} = 0.22$) with directional RandomNormal kernel initialization toward positive values ($\mu = 0.4$, $\sigma = 0.1$) to facilitate spatial separation in the latent space.

Hydrogeomorphological encoder: the hydrogeomorphological encoder processes 18 input variables through four transformation layers, employing an expanded initial configuration to accommodate the greater complexity and spatial heterogeneity of terrain-hydrodynamic interactions. The architectural design prioritizes capturing multi-scale geomorphological patterns and their coupling with hydrodynamic forcing.

$$\mathbf{h}_1^{\text{hydro}} = \text{SELU}(\text{BN}(\mathbf{W}1^{\text{hydro}}\mathbf{x}\text{hydro} + \mathbf{b}_1^{\text{hydro}})), \quad \mathbb{R}^{18} \to \mathbb{R}^{224} \tag{7}$$

$$\mathbf{h}_2^{\text{hydro}} = \text{ELU}(\text{BN}(\mathbf{W}_2^{\text{hydro}}\mathbf{h}_1^{\text{hydro}} + \mathbf{b}_2^{\text{hydro}})), \quad \mathbb{R}^{224} \to \mathbb{R}^{112} \tag{8}$$

$$\mathbf{h}_3^{\text{hydro}} = \text{ReLU}(\text{BN}(\mathbf{W}_3^{\text{hydro}}\mathbf{h}_2^{\text{hydro}} + \mathbf{b}_3^{\text{hydro}})), \quad \mathbb{R}^{112} \to \mathbb{R}^{112} \tag{9}$$

$$\mathbf{h}_4^{\text{hydro}} = \text{SiLU}(\text{BN}(\mathbf{W}_4^{\text{hydro}}\mathbf{h}_3^{\text{hydro}} + \mathbf{b}_4^{\text{hydro}})), \quad \mathbb{R}^{112} \to \mathbb{R}^{64} \tag{10}$$

The hydrogeomorphological latent representation employs the same bounded activation with distinct regularization parameters:

$$\mathbf{z}\text{hydro} = \tanh(\mathbf{W}\text{LE2}\mathbf{h}4^{\text{hydro}} + \mathbf{b}\text{LE2}), \quad \mathbb{R}^{64} \to \mathbb{R}^{1} \tag{11}$$

where dropout rates range from 0.12 to 0.22 across layers, providing more aggressive regularization compared to the biogeochemical branch. The latent layer uses L1-L2 regularization ($\lambda_{L1}$ = 0.10, $\lambda_{L2}$ = 0.13) with directional RandomNormal kernel initialization toward negative values ($\mu$ = -0.6, $\sigma$ = 0.12), promoting orthogonality with the biogeochemical latent dimension.

Latent space representation: the two independent encoding pathways converge into a two-dimensional orthogonal latent space through concatenation:

$$\mathbf{z} = [\mathbf{z}\text{biogeo}, \mathbf{z}\text{hydro}] \in \mathbb{R}^2 \tag{12}$$

This architecture enforces functional disentanglement through the directional kernel initialization described above and explicit orthogonality constraints during training (see Eq. 27), ensuring that biogeochemical and hydrogeomorphological features are independently represented in perpendicular axes of the latent space. This separation enables interpretable decomposition of habitat characteristics into distinct functional components, facilitating ecological interpretation of the compressed representation.

Biogeochemical decoder: the biogeochemical decoder reconstructs the original 15 variables through five expanding layers that mirror the encoder architecture in reverse, progressively transforming the one-dimensional latent representation back to the full variable space. Each layer applies batch normalization before activation to maintain training stability during reconstruction.

$$\mathbf{d}_1^{\text{biogeo}} = \text{SELU}(\text{BN}(\mathbf{W}1^{\text{dec}}\mathbf{z}\text{biogeo} + \mathbf{b}_1^{\text{dec}})), \quad \mathbb{R}^{1} \to \mathbb{R}^{64} \tag{13}$$

$$\mathbf{d}_2^{\text{biogeo}} = \text{GELU}(\text{BN}(\mathbf{W}_2^{\text{dec}}\mathbf{d}_1^{\text{biogeo}} + \mathbf{b}_2^{\text{dec}})), \quad \mathbb{R}^{64} \to \mathbb{R}^{96} \tag{14}$$

$$\mathbf{d}_3^{\text{biogeo}} = \text{ELU}(\text{BN}(\mathbf{W}_3^{\text{dec}}\mathbf{d}_2^{\text{biogeo}} + \mathbf{b}_3^{\text{dec}})), \quad \mathbb{R}^{96} \to \mathbb{R}^{112} \tag{15}$$

$$\mathbf{d}_4^{\text{biogeo}} = \text{SiLU}(\text{BN}(\mathbf{W}_4^{\text{dec}}\mathbf{d}_3^{\text{biogeo}} + \mathbf{b}_4^{\text{dec}})), \quad \mathbb{R}^{112} \to \mathbb{R}^{176} \tag{16}$$

$$\mathbf{d}_5^{\text{biogeo}} = \text{ReLU}(\text{BN}(\mathbf{W}_5^{\text{dec}}\mathbf{d}_4^{\text{biogeo}} + \mathbf{b}_5^{\text{dec}})), \quad \mathbb{R}^{176} \to \mathbb{R}^{256} \tag{17}$$

The final output layer employs sigmoid activation to ensure reconstructed values remain within the normalized [0, 1] range, matching the preprocessing applied to input variables:

$$\hat{\mathbf{x}}\text{biogeo} = \sigma(\mathbf{W}\text{out}^{\text{biogeo}}\mathbf{d}5^{\text{biogeo}} + \mathbf{b}\text{out}^{\text{biogeo}}), \quad \mathbb{R}^{256} \to \mathbb{R}^{15} \tag{18}$$

Hydrogeomorphological decoder: the hydrogeomorphological decoder follows an analogous expansion strategy, reconstructing the 18 original variables from the one-dimensional latent representation through four expanding layers:

$$\mathbf{d}_1^{\text{hydro}} = \text{SiLU}(\text{BN}(\mathbf{W}1^{\text{hdec}}\mathbf{z}\text{hydro} + \mathbf{b}_1^{\text{hdec}})), \quad \mathbb{R}^1 \to \mathbb{R}^{64} \tag{19}$$

$$\mathbf{d}_2^{\text{hydro}} = \text{ReLU}(\text{BN}(\mathbf{W}_2^{\text{hdec}}\mathbf{d}_1^{\text{hydro}} + \mathbf{b}_2^{\text{hdec}})), \quad \mathbb{R}^{64} \to \mathbb{R}^{72} \tag{20}$$

$$\mathbf{d}_3^{\text{hydro}} = \text{ELU}(\text{BN}(\mathbf{W}_3^{\text{hdec}}\mathbf{d}_2^{\text{hydro}} + \mathbf{b}_3^{\text{hdec}})), \quad \mathbb{R}^{72} \to \mathbb{R}^{104} \tag{21}$$

$$\mathbf{d}_4^{\text{hydro}} = \text{SELU}(\text{BN}(\mathbf{W}_4^{\text{hdec}}\mathbf{d}_3^{\text{hydro}} + \mathbf{b}_4^{\text{hdec}})), \quad \mathbb{R}^{104} \to \mathbb{R}^{208} \tag{22}$$

The output layer mirrors the biogeochemical branch with sigmoid-bounded reconstruction:

$$\hat{\mathbf{x}}\text{hydro} = \sigma(\mathbf{W}\text{out}^{\text{hydro}}\mathbf{d}4^{\text{hydro}} + \mathbf{b}\text{out}^{\text{hydro}}), \quad \mathbb{R}^{208} \to \mathbb{R}^{18} \tag{23}$$

where σ(·) denotes the sigmoid activation function, ensuring outputs are bounded in [0, 1] and compatible with the normalized input data distribution.

Training objective: the model is trained by minimizing a composite loss function that balances reconstruction fidelity across both functional domains with explicit enforcement of latent space orthogonality and appropriate regularization:

$$\mathcal{L}\text{total} = \mathcal{L}\text{recon}^{\text{biogeo}} + \mathcal{L}\text{recon}^{\text{hydro}} + \lambda\text{ortho}\mathcal{L}\text{ortho} + \lambda L1|\mathbf{W}\text{LE}|1 + \lambda L2|\mathbf{W}\text{LE}|_2^2 \tag{24}$$

where the reconstruction losses are defined as mean squared error over all training samples:

$$\mathcal{L}\text{recon}^{\text{biogeo}} = \frac{1}{N}\sum i = 1^N|\mathbf{x}\text{biogeo}^{(i)} - \hat{\mathbf{x}}\text{biogeo}^{(i)}|_2^2 \tag{25}$$

$$\mathcal{L}\text{recon}^{\text{hydro}} = \frac{1}{N}\sum i = 1^N|\mathbf{x}\text{hydro}^{(i)} - \hat{\mathbf{x}}\text{hydro}^{(i)}|_2^2 \tag{26}$$

The orthogonality constraint enforces independence between the two latent dimensions by penalizing their absolute dot product across the training batch:

$$\mathcal{L}\text{ortho} = \frac{1}{N}\sum i = 1^N|\mathbf{z}\text{biogeo}^{(i)T}\mathbf{z}\text{hydro}^{(i)}| \tag{27}$$

This term drives the latent representations toward orthogonality, ensuring that biogeochemical and hydrogeomorphological features occupy independent, perpendicular directions in the two-dimensional latent space. When $\mathbf{z}_{\text{biogeo}}$ and $\mathbf{z}_{\text{hydro}}$ are perfectly orthogonal, their dot product equals zero, and Lortho = 0. The regularization terms with $\lambda_{L1}$ = 0.08 (biogeochemical) and 0.10 (hydrogeomorphological), and $\lambda_{L2}$ = 0.22 (biogeochemical) and 0.13 (hydrogeomorphological) prevent overfitting in the latent encoding layers by penalizing large weight magnitudes. The orthogonality weight λortho was optimized through systematic hyperparameter tuning via grid search (see Supplementary Materials).

### 2.3.2 Multi-Objective Loss Function and Optimization Strategy

The multi-objective loss function incorporated six balanced components to optimize the specialization of latent representations, following similar approaches that have demonstrated that multi-objective optimization provides effectiveness in machine learning applications associated with complex multidimensional systems (Xu et al., 2025). The components included: direct specialization through categorical alignment, intensified regularization for the hydrogeomorphological latent representation, spatial separation in latent space, penalization of cross-contamination between categories, controlled orthogonality between representations, and balance of inter-category variances. This architecture prioritized hydrogeomorphological specialization through multiple domain-specific regularization terms (variance, spatial position, activity, and dynamic range).

To address the imbalance observed in specialization of latent representations, a differential weighting of loss functions was implemented, assigning weights 6-18 times greater to hydrogeomorphological reconstruction relative to biogeochemical. This methodological strategy, complemented by the domain-specific regularization terms mentioned earlier, was used to balance representational capacity between both environmental variable categories, allowing the model to effectively capture the distinctive characteristics of each domain, addressing known challenges in training multi-branch autoencoders for heterogeneous environmental data (Heimbach et al., 2025).

The optimization process was structured through random hyperparameter search evaluating 30 architectural configurations with Keras Tuner. During training, specialized callbacks were used for monitoring specialization progress every 4 epochs, automatic memory management, and early stopping with a patience of 55 epochs. The Adam optimizer was configured with adaptive logarithmic learning rate ($5\times10^{-5}$ to $8\times10^{-4}$) and gradient clipping (norm=0.2) to ensure stability during the optimization process.

### 2.3.3 Training Protocol and Convergence

Model training was executed for 220 epochs using a stratified train/validation partition of 80:20 with fixed seed (seed=42) to ensure reproducibility of results. During this process, specialized callbacks were implemented that included: early stopping (EarlyStopping) with validation loss monitoring and patience of 55 epochs, adaptive learning rate reduction (ReduceLROnPlateau) with reduction factor 0.2 every 30 epochs without improvement, and a custom callback (OptimizedDualCallback) for domain-specific tracking of hydrogeomorphological latent representation metrics every 4 epochs.

This training protocol allowed extraction of two latent variables with functionally differentiated characteristics. The first representation (LE1) captured biogeochemical information compressed with initialization directed toward positive values, while the second (LE2) synthesized hydrogeomorphological characteristics with initialization toward negative values. This architecture achieved effective dimensional reduction from 33 original variables to 2 specialized latent representations, whose convergence was evaluated through joint minimization of reconstruction loss (MSE) and the multi-objective regularization terms previously described (Figure 2).

### 2.3.4 Quality Evaluation of Latent Representations

To quantify the quality of obtained latent representations, an analysis based on the Maximal Information Coefficient (MIC) was implemented using the R minerva package with optimized parameters ($\alpha$=0.6, C=15). This methodology enabled detection of complex nonlinear relationships between latent variables and original environmental variables, surpassing limitations of traditional correlation analyses and providing robust measures of statistical dependence in complex marine systems (Salgado et al., 2024). The calculated MIC matrices incorporated directional information through Spearman correlation, providing a robust measure of the functional specialization of each latent representation.

Evaluation criteria were structured into four complementary metrics: intra-category purity (average MIC between each latent variable and variables of its target domain), inter-category contamination (average MIC with variables from non-target domains), specialization ratio (intra_purity/inter_contamination), and global orthogonality between LE1 and LE2.

## 2.4 Potential Classification of Benthic Habitats

### 2.4.1 Hyperparameter Optimization and SOM Training

The latent features LE1 and LE2 obtained from the autoencoder were normalized using Z-score transformation to ensure equivalent contributions during unsupervised classification. This preprocessing proved essential for correct functioning of the SOM (Self-Organizing Map) algorithm, as it reduced biases derived from the different scales of latent representations. Self-organizing maps were used for identification of potential benthic habitats because they have demonstrated particular effectiveness in marine ecology applications, particularly for identification of community patterns and benthic habitat classification, preserving topological relationships in multidimensional environmental data (Chon et al., 2023; Song et al., 2007).

Hyperparameter optimization was executed through exhaustive search systematically evaluating multiple architectural and training configurations. Rectangular grid topologies evaluated included dimensions ranging from 1×1 to 20×20 neurons (1×1, 4×4, 5×5, 7×7, 10×10, and 20×20), while training parameters encompassed 50, 100, and 200 epochs with decreasing learning rates varying from initial-final α configurations of (1.0, 0.5), (1.0, 0.1), (1.0, 0.01), (0.5, 1.0), (0.5, 0.5), and (1.0, 0.1), and neighborhood radii with gradual reduction of (10, 5) and (5, 1). This methodological approach enabled identification of the configuration that jointly minimized quantization error (average Euclidean distance between input vectors and winning neurons) and topographic error (proportion of vectors whose winning neuron and second-best neuron are not topological neighbors) (Liu et al., 2006).

### 2.4.2 Generation of Potential Benthic Habitat Map

The potential benthic habitat map was obtained through conversion of geographic coordinates and habitat classes to raster format using R spatial functions (Hijmans, 2023). The resulting raster map maintained a spatial resolution of 0.04°, consistent with the resolution of the original environmental variables, and was exported in GeoTIFF format for compatibility with standard Geographic Information System applications. Each georeferenced pixel represented a specific benthic habitat class characterized by its unique biogeochemical and hydrogeomorphological signature, providing a spatially explicit basis for subsequent ecological analyses and adaptive marine management applications.

## 2.5 Validation of Potential Benthic Habitat Classification

### 2.5.1 Environmental Variability Patterns Captured by Latent Space

Validation of the autoencoder's semantic compression capacity was performed through analysis of correspondence between the two-dimensional latent space and the original environmental variables. Specifically, the two compressed latent variables (LE1 and LE2) were related to each of the 33 environmental variables through scatter plots, ensuring beforehand that all environmental layers were spatially aligned and rescaled to a homogeneous resolution of 0.04°. To facilitate ecological interpretation of learned representations, each point in the scatter plots was labeled according to the class assigned by the SOM model, thus enabling visualization of cluster distribution in latent space. This analytical approach allowed corroboration that the clustering structure captured by the autoencoder preserved coherence with known environmental variability patterns of the Colombian Pacific Ocean, validating that compression not only reduced dimensionality efficiently, but also retained ecologically interpretable information spatially consistent with regional oceanographic dynamics (Figure 2).

#### 2.5.2 Ecological Validation Through Species Occurrence Records

##### 2.5.2.1. Species Occurrence Data and Quality Control

Occurrence records of multiple marine species associated with the study area were compiled from specialized institutional databases focusing on marine biodiversity, including AUNAP (National Authority for Aquaculture and Fisheries) (https://aunap.gov.co/), INVEMAR (Institute for Marine and Coastal Research, 2023) (https://www.invemar.org.co/), GBIF (Global Biodiversity Information Facility) (https://www.gbif.org/), and OBIS (Ocean Biodiversity Information System) (https://obis.org/). Rigorous quality filters were applied following standard protocols for marine biodiversity data: elimination of missing or implausible coordinates, duplicate records and georeferencing errors, as have been implemented by other studies with similar approaches (Navarro et al., 2023). Records were tabulated in structured format (longitude, latitude, kingdom, phylum, class, order, family, genus, scientific name, and source) and classified into two habitat categories: benthic (with relationships to the seafloor) and purely pelagic (with little or no relationship to the seafloor).

##### 2.5.2.2. Hierarchical Species Filtering and Habitat Preference Classification

A sequential filtering process was applied to identify species with robust biological representation, following established criteria for marine species distribution studies. Initially, only species with a minimum of ten independent occurrence records were retained, a threshold that ensures sufficient statistical power to characterize habitat preferences (Guisan & Thuiller, 2005; Naeem et al., 2016). Subsequently, filtered species were classified into two mutually exclusive habitat categories: benthic (with documented ecological relationship to marine substrate) and purely pelagic (restricted to the water column throughout their entire life cycle). Finally, niche specialization degree was quantified using Levins standardized niche breadth index ($B_{std} = B/n$, where $B = 1/\Sigma p_i^2$), classifying species into three categories: specialists ($B_{std} < 0.3$)—associated with restrictive environments; selective ($0.3 \leq B_{std} < 0.6$)—with moderate preferences; generalists ($B_{std} \geq 0.6$)—tolerant of wide environmental ranges (Devictor et al., 2008; Julliard et al., 2006). This hierarchical approach enables validation of SOM classification coherence across ecologically relevant subsets, avoiding biases derived from dominant species or particular habitat categories.

##### 2.5.2.3. Correspondence Between Environmental Gradients and Species Composition Patterns

Filtered species according to abundance, habitat, and specialization criteria were used to evaluate correspondence between environmental gradients and species composition patterns in the potential habitat map derived from the SOM. Pairwise Mantel test was implemented comparing three environmental distance metrics calculated in integrated latent space (LE1 and LE2): Euclidean, Manhattan, and Mahalanobis distance. Each environmental distance matrix was paired against the species composition dissimilarity matrix (Bray-Curtis index) through Pearson correlation under 9,999 random permutations, following robust statistical protocols for correspondence analysis in marine ecology (Legendre & Legendre, 2012). The distance method presenting the highest correlation coefficient (r) was selected as indicator of the optimal metric for explaining observed biological patterns. Result interpretation followed established criteria: $r > 0.5$ strong correlation, $0.3 < r \leq 0.5$ moderate, $r \leq 0.3$ weak, considering statistical significance at $p < 0.05$. This analysis validated whether SOM-based zonation captured environmental gradients with predictive power over marine species distribution (Clarke et al., 2014).

#### 2.5.2.4 Environmental Correspondence Between Potential Habitats and Species Environmental Preferences

Community differentiation among clusters was quantified using the Bray-Curtis dissimilarity index (range 0-1), which quantified species turnover (β-diversity) among cluster pairs. For this test, values close to 1 indicate high community differentiation, while values close to 0 indicate biological homogeneity (Baselga, 2010). To identify characteristic species of each cluster, the Indicator Value Index (IndVal) was calculated, which integrates two complementary components: (1) specificity (A), defined as the proportion of occurrences within the cluster that correspond to a given species, quantifying the degree of preferential species association with that cluster; and (2) fidelity (B), defined as the proportion of total species occurrences found within the cluster, reflecting the constancy of its presence in that group (De Cáceres et al., 2012). IndVal was calculated as $IndVal = \sqrt{(A \times B)} \times 100$, where values range between 0 and 100; high values (≥70) indicate that the species presents both high specificity and fidelity simultaneously, establishing it as a robust indicator of the particular environmental conditions of the cluster, with only species showing statistically significant IndVal considered.

In the event that high differentiation among clusters observed through β-diversity (Bray-Curtis dissimilarity values ≥0.7) generates communities with high specificity but low fidelity—characterized by species with distributions restricted to particular clusters but with insufficient absolute abundances—IndVal analysis might not identify an adequate number of statistically significant indicator species ($p \leq 0.05$) to exhaustively characterize delimited habitats. Facing this potential limitation, an alternative method is proposed: analysis of dominant species composition by cluster through dual thresholds: ≥10% relative composition (specificity A) and a minimum of 10 independent occurrences per species (González-Valdivia et al., 2011). This complementary approach enables: (1) identification of ecologically relevant species in clusters with low record density, (2) extraction of observed environmental ranges (temperature, salinity, depth) by cluster from oceanographic variables, and (3) systematic contrast of these ranges against documented tolerance intervals in scientific literature for dominant species, thus validating the biogeographic coherence of habitat classification and detecting possible incongruences derived from limitations in environmental variables or niche specificities not captured by the two-dimensional latent space.

## 3. Results and Discussion

### 1.1 Architecture and Training of the Autoencoder (ECOSAIC)

#### 1.1.1 Model Convergence and Multi-Objective Loss Function Performance

ECOSAIC model training over 220 epochs demonstrated successful convergence of the multi-objective loss function, achieving stabilization after approximately 50 epochs (Figure 3-A). This convergence behavior is consistent with patterns observed in successful multi-objective autoencoder applications for complex systems, where early stabilization indicates well-balanced architecture and appropriate optimization parameters (Xu et al., 2025). Total loss exhibited exponential reduction from initial values exceeding 11.0 to stabilization around 0.4, with parallel trajectories between training and validation indicating absence of overfitting (Üstek et al., 2024). The ratio of validation losses between hydrogeomorphological and biogeochemical components (Figure 3-B) stabilized around one (1.0) after the first 50 epochs, with fluctuations between 0.6 and 1.5 throughout training. This behavior indicated that both latent components (LE1 and LE2) achieved comparable levels of reconstruction error, suggesting equitable balance in the model's representational capacity. Convergence toward values close to unity implies that, regardless of the differential weighting implemented in the loss function, the dual autoencoder architecture with orthogonal regularization favors balanced error distribution between the two functional categories.

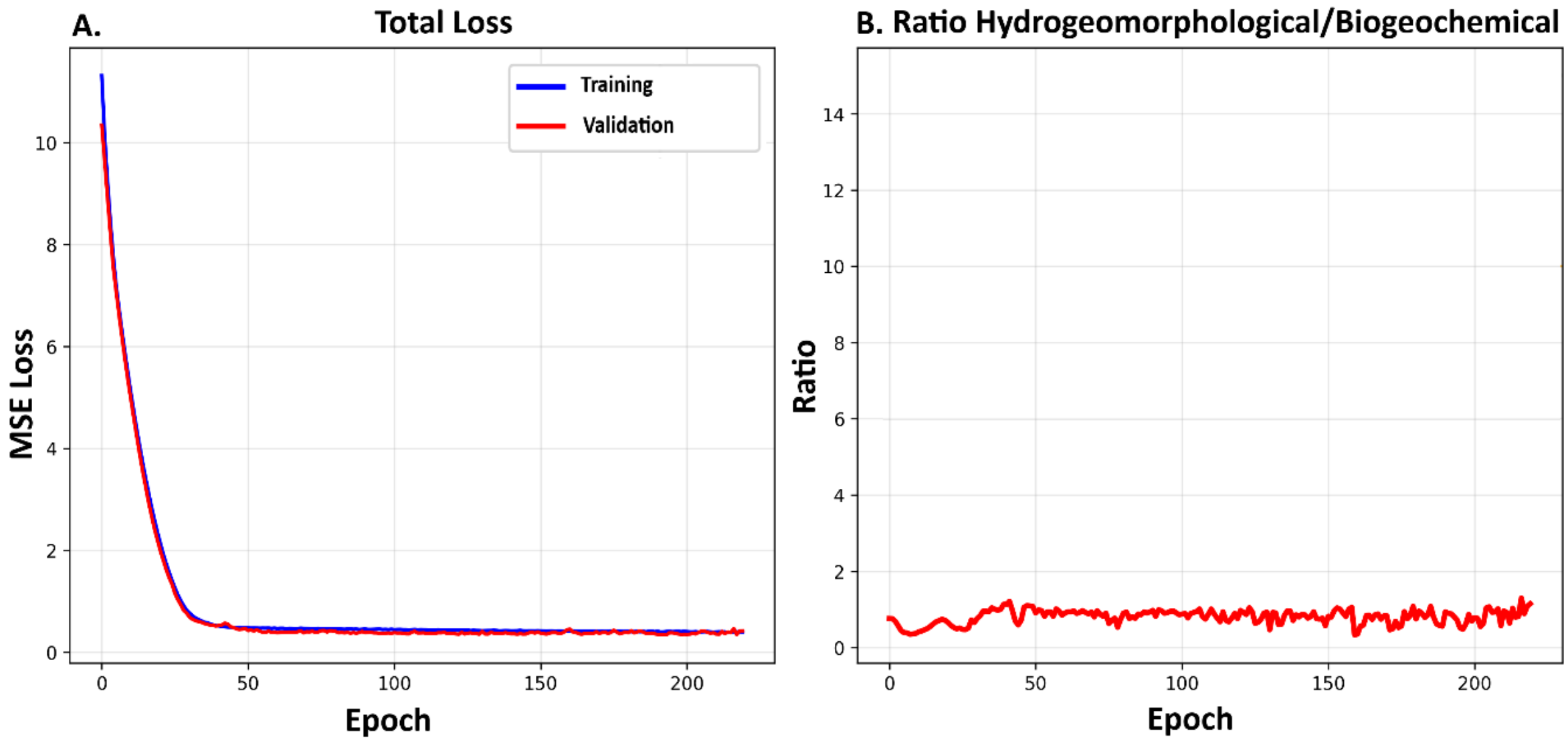

Figure 3. Training convergence (A) and reconstruction loss balance (B) for the ECOSAIC model.

#### 1.1.2 Extraction and Characteristics of Latent Representation

The ECOSAIC model extracted two one-dimensional latent representations (LE1 and LE2). LE1, derived from compression of the upper autoencoder branch, compressed 15 variables, capturing physicochemical gradients and biological productivity characteristic of the seafloor in the Colombian Pacific Ocean. LE2, resulting from compression of the lower autoencoder branch, compressed 18 variables, capturing the geomorphological structure of the substrate and benthic hydrodynamic processes of the study area. Dimensional compression from 33 to 2 variables (93.9%) preserved ecologically relevant information, a result that compares favorably with previous studies of dimensional reduction in oceanographic data that typically achieve high compression ratios while maintaining functional information (Baig et al., 2023; Falasca et al., 2024).

### 1.2 Quality and Specialization of Latent Space

#### 1.2.1 Categorical Purity Analysis Using Maximal Information Coefficient (MIC)

Quantitative analysis of specialization revealed marked differences between both latent components in terms of functional purity across domains. LE1 exhibited robust specialization toward its target biogeochemical domain, with an average intra-category MIC of 0.713 versus inter-category contamination of 0.391 (specialization ratio = 1.83), indicating moderately strong functional segregation across dimensions. In contrast, LE2 showed limited specialization toward its target hydrogeomorphological domain (intra-category MIC = 0.436) with comparable inter-category contamination (MIC = 0.411), resulting in a specialization ratio close to unity (1.06). This asymmetry in specialization patterns suggests that, while LE1 effectively captures a differentiated biogeochemical axis, LE2 operates in a region of latent space where hydrogeomorphological and biogeochemical signals exhibit tighter coupling.

LE1 demonstrated specialization toward its target biogeochemical domain, exhibiting strong positive MIC correlations with eight variables of this category: salinity (0.978), nitrate (0.888), phosphate (0.873), pH (0.536), and utilized oxygen (0.482), together with five hydrogeomorphological variables with lower MIC values. In the negative range, LE1 showed 11 correlations with biogeochemical variables, highlighting net primary productivity (-0.985), mean temperature (-0.943), phytoplankton (-0.942), bottom light (-0.821), and dissolved oxygen (-0.810), plus eight hydrogeomorphological variables with minor MIC contribution. This polarized structure suggests that LE1 captured a biogeochemical gradient where availability of inorganic nutrients opposes active photosynthetic activity (Figure 4).

LE2 showed limited specialization toward its hydrogeomorphological domain, with positive correlations in eight variables of this category: Radial Influence of Mangrove Biomass (IRBM) (0.758), light attenuation coefficient (0.667), euphotic zone depth (0.655), longitudinal river influence (0.663), grain size (0.547), bathymetric standard deviation (0.455), slope (0.432), and runoff (0.337), accompanied by ten biogeochemical variables with lower MIC values. Negative correlations included seven hydrogeomorphological variables: wind velocity (-0.722), local bathymetric mean (-0.604), distance to coast (-0.600), bathymetric surface (-0.601), northward talus component (-0.364), current velocity (-0.354), and bathymetric aspect (-0.339), together with eight biogeochemical variables with minor MIC contribution (Figure 4). This structuring reflects a coast-ocean gradient where terrigenous influence of mangroves and rivers contrasts with deep oceanic conditions and exposed areas, a pattern widely documented in tropical coastal systems of the Eastern Pacific (Gómez & Bernal, 2013)

Despite regularization efforts aimed at promoting functional specialization for LE1 and LE2, both latent components exhibited cross-contamination (Figure 4). LE1 showed moderately strong MIC values with hydrogeomorphological variables, particularly with mean depth (0.741), bathymetric surface (0.682), and distance to coast (0.442) in the positive range, and with IRBM (-0.614), euphotic zone depth (-0.599), and light attenuation coefficient (-0.592) in the negative range. Similarly, LE2 correlated with biogeochemical variables outside its target domain: phytoplankton (0.610), light at bottom (0.584), primary productivity (0.575), mean temperature (0.567), and chlorophyll-a (0.556) in the positive range, along with nitrate (-0.554), salinity (-0.552), and phosphate (-0.533) in the negative range. These findings suggest that the biogeochemical and hydrogeomorphological dimensions are intrinsically coupled in benthic ecosystems of the Eastern Tropical Pacific, where processes such as light availability (determined by depth and turbidity) and nutrient resuspension (mediated by hydrodynamics) functionally link both domains, limiting their complete orthogonality in latent space. This interconnection is characteristic of tropical coastal systems where bathymetric gradients, terrigenous influence, and biogeochemical processes interact in a complex manner.

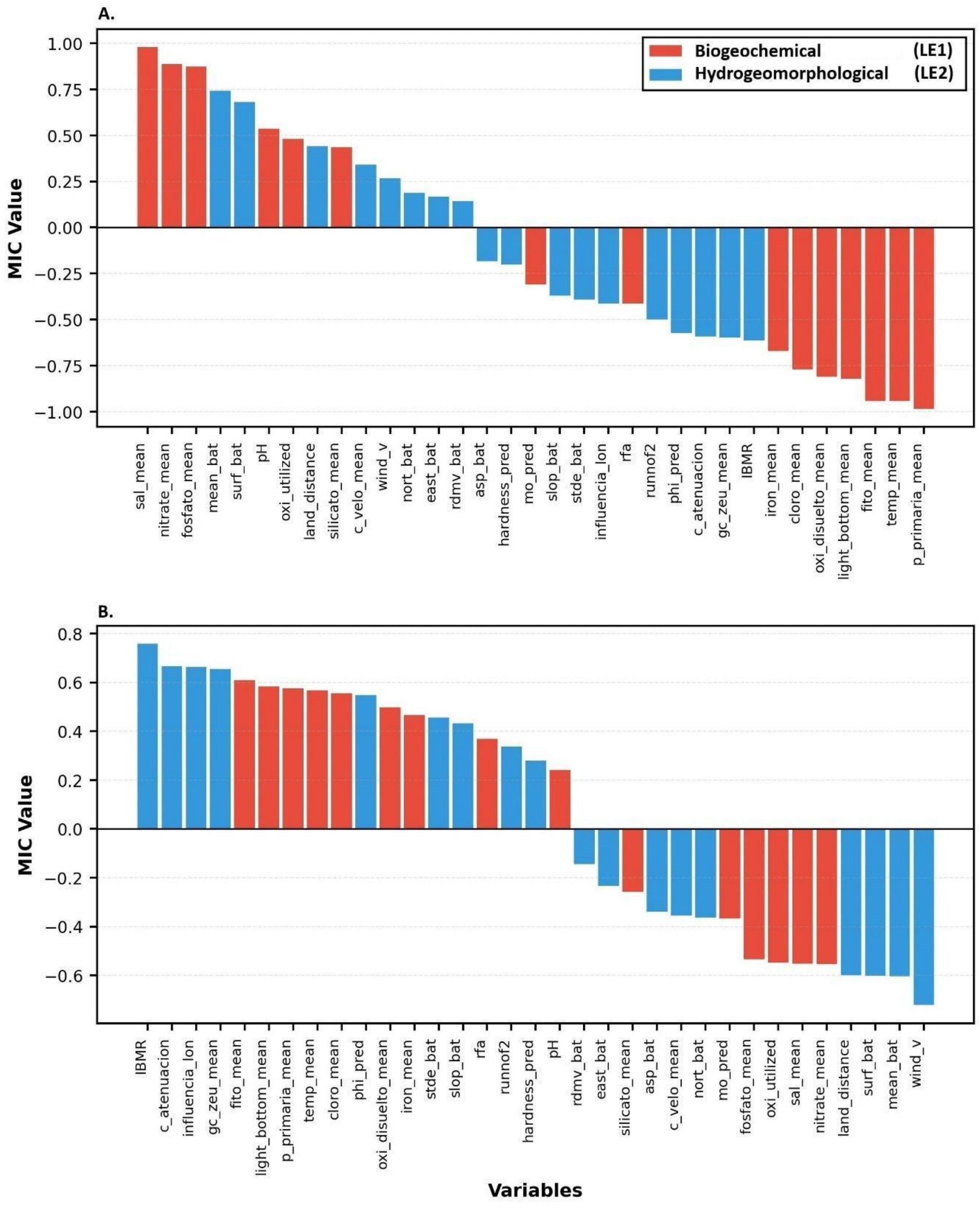

Figure 4. Functional specialization of latent components LE1 (A) and LE2 (B) evaluated through Maximum Information Coefficient (MIC) correlations with biogeochemical variables (red bars) and hydrogeomorphological variables (blue bars).

### 1.2.2 Orthogonality Between Latent Features

The orthogonality analysis revealed substantial linear separation between latent representations (Pearson $r = -0.276$, $p < 0.001$; $R^2 = 0.076$), with only 7.6% of variance linearly shared between LE1 and LE2 ($n = 1223$; Figure 5). This linear orthogonality confirms the success of the dual architecture in generating functionally differentiated latent dimensions, consistent with the spatially separated distributions observed (LE1: $\mu = -0.003 \pm 0.041$; LE2: $\mu = -0.079 \pm 0.109$). The degree of orthogonality achieved compares favorably with previous studies of specialized autoencoders applied to multivariate environmental data, where typical cross-correlation values range between 0.15–0.35 (Baig et al., 2023). However, analysis using Maximum Information Coefficient (MIC) revealed moderate nonlinear dependence (MIC = −0.607), suggesting possible residual functional coupling between the biogeochemical and hydrogeomorphological domains. This nonlinear coupling may reflect intrinsic ecological interactions in the Eastern Tropical Pacific, where processes such as light attenuation (determined by depth and turbidity) and nutrient resuspension (mediated by benthic hydrodynamics) naturally link both domains, limiting their complete separability in latent space. The presence of residual nonlinear dependencies is consistent with current understanding of coastal marine systems, where inherent ecological complexity prevents complete orthogonality of functional domains (Falasca et al., 2024).

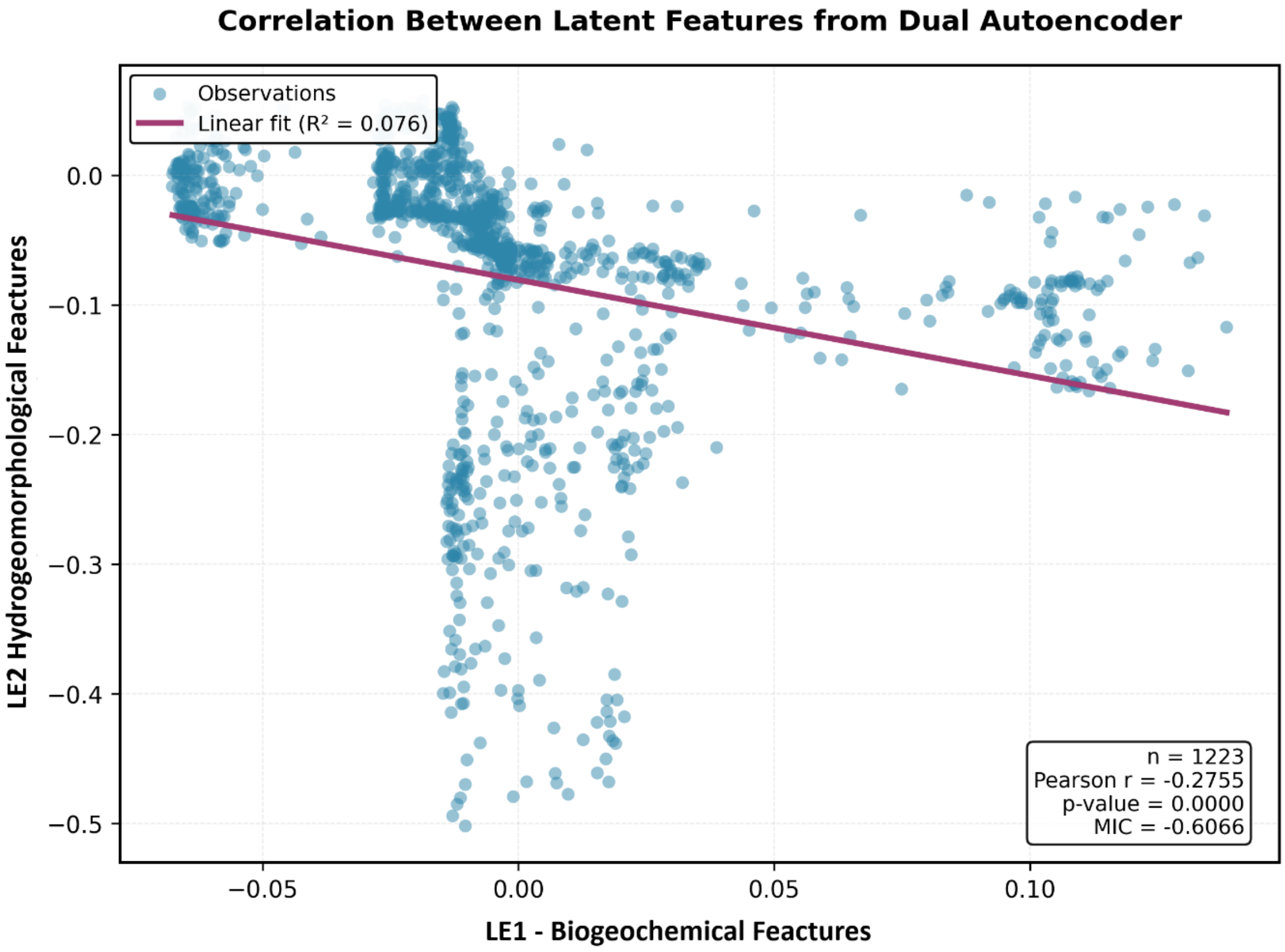

Figure 5. Evaluation of orthogonality between latent representations (LE1 and LE2) obtained using the ECOSAIC model.

The optimal SOM configuration consisted of a 4×4 rectangular grid trained for 100 epochs with decreasing learning rate (α: 1.0→0.01) and neighborhood radius reducing from 5 to 1. This parametrization achieved quantization error of 0.281 and topographic error of 0.026, ensuring precision in class assignment and spatial coherence of the resulting map. The error values obtained fall within ranges considered optimal for SOM applications in marine habitat classification, where quantization errors below 0.3 and topographic errors below 0.05 indicate satisfactory convergence and preservation of spatial relationships (Chon et al., 2023; Song et al., 2007)

SOM analysis identified 16 benthic habitat classes distributed across approximately 20,049 km² of the study area, revealing a marked dichotomy between coastal and oceanic habitats (Figure 6). Coastal habitats, especially those located in the southern and central Colombian Pacific zone with high influence of mangroves and sediment discharge from rivers (clusters 5, 1, and 9), which occupied approximately 5,619 $km^2$ (28% of total area), were characterized by distinctive biogeochemical conditions associated with high primary productivity and strong terrigenous influence. These habitats presented elevated temperatures (mean: 26.01 ± 2.04 °C), high concentrations of net primary productivity (0.0247 ± 0.0075 [$g \cdot m^{-3} \cdot day^{-1}$]) and phytoplankton (1.92 ± 0.38 [$\mu mol \cdot m^{-3}$]), abundant bottom light (18.70 ± 10.04), and elevated values associated with IRBM (0.661 ± 0.166) that evidenced proximity to complex coastal ecosystems. This characterization is consistent with previous studies documenting the significant influence of mangroves on surrounding benthic communities in the Colombian Pacific, modifying physicochemical characteristics of sediment and providing critical ecological connectivity (Alfaro & Alfaro, 2010; Gómez & Bernal, 2013).

In contrast, deep oceanic habitats (clusters 11, 12, 14, 15, and 16) encompassed approximately 7,288 $km^2$ (36% of total area), displaying characteristics of stratified waters far from the coast: elevated salinities (34.79 ± 0.17 [PSU]), high concentrations of nitrate (26.86 ± 7.38 [$mol \cdot m^{-3}$]) and phosphate (2.11 ± 0.53 [$mol \cdot m^{-3}$]), greater bathymetric depth (-141.2 ± 136.7 m), considerable distances from coast (0.265 ± 0.084 [degrees]), and intense wind forcing exposure (wind velocity: 0.094 ± 0.030 [$m \cdot s^{-1}$]). Notably, the two oceanic clusters of greatest extent (14 and 16) concentrated approximately 5,372 $km^2$ (27% of total area), representing the most extensive benthic habitats, characterized by deep oligotrophic conditions with limited surface primary production, yet enrichment of inorganic nutrients through benthic remineralization processes. These characteristics are typical of Eastern Tropical Pacific waters far from continental influence, where water column stratification and deep upwelling processes determine nutrient distribution.

Unlike habitats located in the southern and central Colombian Pacific zone, northern habitats, especially those identified as clusters 8, 3, and 4, which represented approximately 3,302 $km^2$ (16% of total area), exhibited distinctive characteristics possibly associated with equatorial upwelling system influence (Figure 6). These habitats presented lower temperatures (19.05 ± 4.78°C) indicative of subsurface water upwelling, moderate primary productivity (0.0084 ± 0.0066 [$g \cdot m^{-3} \cdot day^{-1}$]), moderate phytoplankton concentrations (0.85 ± 0.51 [$\mu mol \cdot m^{-3}$]), and lower IRBM values (0.198 ± 0.090) reflecting reduced mangrove ecosystem influence and more exposed oceanic conditions. This latitudinal differentiation suggests that central-southern benthic habitats are structured primarily by continental forcing associated with massive river discharge and terrigenous inputs, while northern habitats respond predominantly to mesoscale oceanographic forcing (coastal upwelling), configuring two ecologically distinct regions in the Colombian Pacific coastal continuum. The influence of upwelling processes on Eastern Tropical Pacific benthic habitat structuring has been previously documented, where variations in temperature and nutrient availability associated with these processes determine species distribution patterns and community structure (Spring & Williams, 2023).

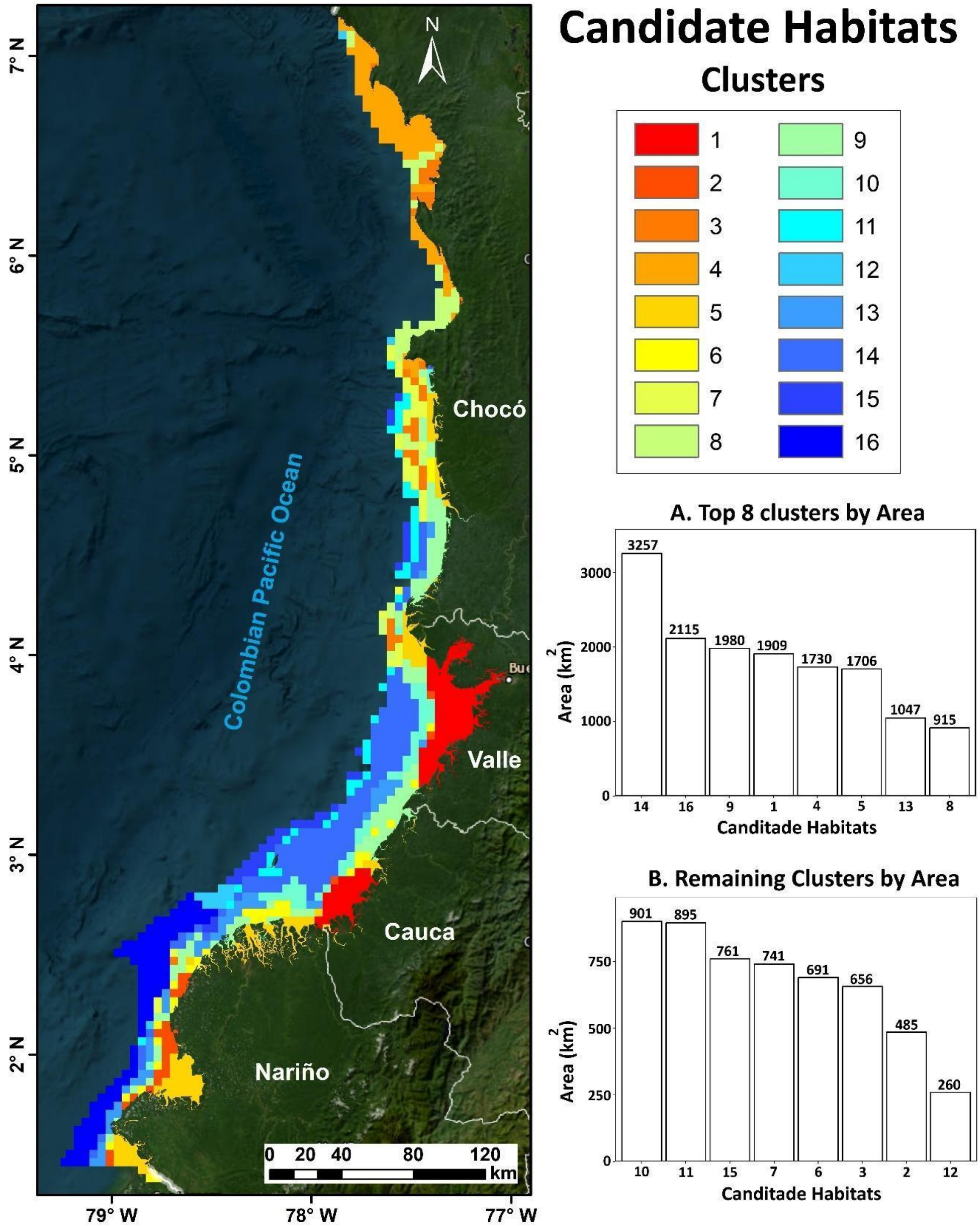

Figure 6. Potential habitat map of the Colombian Pacific Ocean identified using Self-Organizing Map. The figure shows the spatial distribution and area quantification (km²) of each of the 16 potential benthic habitats.

### 1.3 Validation of Potential Benthic Habitat Classification

#### 3.5.1 Environmental Information Preservation in Latent Space

Biogeochemical variables exhibited predominant associations with LE1, confirming that this latent representation effectively captured fundamental environmental gradients for benthic habitat structuring (Figure S1). Salinity (MIC: +0.978) showed a strong positive relationship with ascending linear trend along LE1, reflecting the transition from low concentrations in zones with estuarine influences to high concentrations in oceanic zones far from the coast. Nitrate (MIC: +0.888) and phosphate (MIC: +0.873) exhibited similar patterns of sustained increase with increasing LE1 values, indicating that this representation encodes an inorganic nutrient availability gradient associated with upwelling processes and benthic remineralization in deep waters. These coherent responses reflect LE1's capacity to integrate biogeochemical conditions of the water-sediment interface that determine nutrient availability for benthic communities, patterns that are fundamental in Eastern Tropical Pacific marine ecosystem structuring (Assis et al., 2024).

In contrast, variables negatively associated with LE1 captured biogeochemical conditions related to high photosynthetic activity and biological productivity (Figure S1). Net primary productivity (MIC: -0.985) and phytoplankton concentration (MIC: -0.942) showed pronounced inverse relationships, progressively decreasing toward high LE1 values and defining zones of intense primary production in surface waters close to the coast associated with low LE1 values. Temperature (MIC: -0.943) exhibited a pattern of progressive decrease along LE1, characteristic of thermal stratification from shallow coastal bottoms toward deep waters, while bottom light (MIC: -0.821) and dissolved oxygen (MIC: -0.810) presented concordant decreases associated with bathymetric deepening and reduction in light penetration. This biogeochemical polarization evidences that LE1 functions as a coast-ocean axis capturing the transition from shallow, warm, and biologically productive environments toward deep, cold, and inorganic nutrient-enriched environments, reflecting fundamental oceanographic gradients of the Colombian Pacific (Figure 6).

Hydrogeomorphological variables showed differential structuring along LE2, albeit with lower specialization than observed in LE1 relative to biogeochemical variables (Figure S2). The Radial Index of Mangrove Biomass (IRBM) (MIC: +0.758) and light attenuation coefficient (MIC: +0.667) exhibited positive associations with increases toward high LE2 values, indicating that this representation captures the influence of terrigenous inputs and turbidity associated with mangrove ecosystems and river discharges. The euphotic zone depth (MIC: +0.655) and river length influence (MIC: +0.663) showed patterns of consistent increase with increasing LE2 values, characteristic of shallow coastal zones under strong continental impact where light penetration is limited by suspended particulate material. These responses suggest that high LE2 values represent coastal environments with high ecotonal mangrove-estuarine complexity (Figure 6), consistent with the geomorphological configuration of the Colombian Pacific where mangroves exert significant influence on surrounding benthic habitat characteristics (Gómez & Bernal, 2013).

Variables negatively associated with LE2 concentrated deep oceanic and exposed characteristics (Figure S2). Wind velocity (MIC: -0.722) showed pronounced negative correlation, increasing toward low LE2 values, coherent with greater exposure of zones far from the coast to atmospheric forcing. Local bathymetric mean (MIC: -0.604), distance to coast (MIC: -0.600), and bathymetric surface (MIC: -0.601) exhibited concordant patterns of increase toward low LE2 values, defining a coast-ocean and shallow-deep gradient. Bathymetric slope and roughness variables showed more complex distributions, reflecting geomorphological heterogeneity in transition environments. This structuring confirms that LE2 functions as a secondary axis complementing LE1, capturing primarily hydrogeomorphological characteristics related to the transition from coastal environments protected under terrigenous influence toward exposed and deep oceanic platforms.

### 3.5.2 Ecological Validation Through Species Occurrence

### 3.5.1 Species Filtering and Niche Specialization Classification

From the initial set of 629 species (613 benthic and 16 purely pelagic), the hierarchical filtering applied retained 169 species that met criteria for robust biological representation (minimum of 10 independent occurrences). Classification by niche specialization through the standardized Levins index identified 23 specialist species ($B_{std} < 0.3$) and 146 selective species ($0.3 \leq B_{std} < 0.6$), reflecting a community dominated by species with moderate to restrictive environmental preferences, a pattern consistent with tropical coastal marine ecosystems subjected to pronounced environmental gradients (Devictor et al., 2008; Julliard et al., 2006). This specialization distribution is characteristic of Eastern Tropical Pacific systems, where considerable environmental heterogeneity favors ecological niche differentiation and coexistence of species with specific environmental requirements.

#### 3.5.2 Environmental Gradient Correspondence with Species Composition

Correspondence analysis through Mantel test revealed a strong and significant correlation between environmental distances in the integrated latent space (LE1-LE2) and species composition dissimilarity (Bray-Curtis). The optimal distance metric presented a correlation coefficient of $r = 0.448$ ($p < 0.004$), exceeding the moderate correlation threshold ($r > 0.3$) established by Legendre (2012). This result indicated that SOM-based zonation effectively captured environmental gradients with predictive power over spatial distribution of benthic marine species, validating the ecological coherence of the generated habitat classification. The magnitude of the observed correlation compares favorably with previous studies of benthic habitat validation using species occurrence data, where correlations between approximately 0.35–0.55 are considered indicative of ecologically significant classifications (Legendre & Legendre, 2012).

#### 3.5.3 Community Differentiation Among Potential Habitats

Beta-diversity analysis revealed elevated community differentiation among clusters, with average Bray-Curtis dissimilarity values of 0.826 ± 0.159 (median = 0.887, range = 0.357–1.000). These values close to 1 indicate marked species turnover among potential habitats (Baselga, 2010), suggesting that each cluster represents distinctive biological units with specific species compositions. The moderate variability (SD = 0.159) reflects the existence of both abrupt transitions and smoother gradients among certain habitat pairs, coherent with environmental heterogeneity characteristic of the Colombian Pacific. The levels of community differentiation observed are consistent with patterns reported for tropical benthic ecosystems, where the combination of bathymetric gradients, variable terrigenous influence, and geomorphological heterogeneity promote differentiation of specialized communities.

#### 3.5.4 Species-Habitat Environmental Preference Validation

IndVal analysis revealed a characteristic pattern of high specificity combined with low fidelity in most species analyzed (Table 3). This result indicated that, although species presented distributions concentrated in specific clusters, their relative occurrences within those clusters were consistently low. This community structure, characterized by aggregated but numerically limited distributions, resulted in an insufficient number of statistically significant indicator species ($p \leq 0.05$) to exhaustively characterize all habitats generated through SOM classification.

Analysis based on dominant species, defined as those representing ≥10% of occurrences per cluster with a minimum of 10 independent records, enabled identification of 16 dominant species distributed across 11 of the 16 generated clusters (Table 2), enabling biological validation of approximately 70% of the classified habitats through comparison between characteristic environmental ranges of each cluster and ecological preferences documented in specialized scientific literature. Four clusters (11, 14, 15, and 16) did not strictly meet established criteria; however, a decision was made to include the species with greatest indicator potential (Pocillopora damicornis for cluster 11, Rhizoprionodon longurio for cluster 14, Lepidochelys olivacea for cluster 15, and Tripos fusus for cluster 16), resulting in consistent validation between cluster environmental ranges and ecological preferences of these species. Cluster 12 could not be validated because it presented only 2 intercepted occurrences, representing a knowledge gap requiring additional sampling efforts to complete biological validation of the potential habitat map generated in this study.

Table 3. Biological validation of potential benthic habitats through environmental preferences of dominant species in the Colombian Pacific Ocean. The table integrates number of occurrences (n), specificity (S), and fidelity (F).

| Cluster | Specie | n | S | F | IndVal | Cluster range | Literature range | Units | Validation |
|---|---|---|---|---|---|---|---|---|---|
| 1 | *Pomadasys panamensis* | 345 | 20.31 | 36.7 | 27.3 | [22.69-27.37] | [20-30] | [°C] | Very High |
| | | | | | | [31.76-33.39] | [25-40] | [PSU] | |
| | | | | | | [0-23.89] | [0-70] | [m] | |
| 2 | *Mugil cephalus* | 13 | 15.48 | 54.17 | 28.96 | [22.06-27.16] | [5-30] | [°C) | Very High |
| | | | | | | [32.73-33.85] | [10-40] | [PSU] | |
| | | | | | | [0-39.11] | [0-300] | [m] | |
| 3 | *Lutjanus guttatus* | 48 | 14.59 | 3.8 | 7.45 | [16.15-27.1] | [15-30] | [°C) | Very High |
| | | | | | | [31.85-34.84] | [25-40] | [PSU] | |
| | | | | | | [0-201.25] | [0-170] | [m] | |
| 4 | *Solenocera agassizii* | 80 | 14.13 | 32.92 | 21.57 | [11.63-27.41] | [20-30] | [°C) | High |
| | | | | | | [31.16-34.91] | [25-35] | [PSU] | |
| | | | | | | [0-218.71] | [20-500] | [m] | |
| 5 | *Litopenaeus occidentalis* | 271 | 44.87 | 28.83 | 35.97 | [22.72-27.4] | [21.3-28.6] | [°C) | Very High |
| | | | | | | [31.51-33.78] | [25-35] | [PSU] | |
| | | | | | | [0-29.89] | [2-160] | [m] | |
| 6 | *Litopenaeus occidentalis* | 143 | 36.29 | 15.21 | 23.49 | [20.38-27.21] | [21.3-28.6] | [°C) | Very High |
| | | | | | | [31.84-33.97] | [25-35] | [PSU] | |
| | | | | | | [0-71.56] | [2-160] | [m] | |
| 7 | *Xiphopenaeus riveti* | 32 | 14.48 | 13.5 | 13.98 | [8.18-26.81] | [10-30] | [°C) | High |
| | | | | | | [33.25-34.96] | [25-35] | [PSU] | |
| | | | | | | [8-345.33] | [2-80] | [m] | |
| 8 | *Penaeus brevirostris* | 248 | 26.11 | 76.78 | 44.77 | [7.33-25.62] | [15-27] | [°C) | High |
| | | | | | | [33.41-34.95] | [25-35] | [PSU] | |
| | | | | | | [0-218.67] | [10-450] | [m] | |
| 9 | *Litopenaeus occidentalis* | 251 | 21.33 | 26.7 | 23.86 | [14.55-27.31] | [21.3-28.6] | [°C) | High |
| | | | | | | [31.92-34.95] | [25-35] | [PSU] | |
| | | | | | | [0-102.44] | [2-160] | [m] | |
| 10 | *Pomadasys panamensis* | 126 | 24.51 | 13.4 | 18.12 | [18.86-27.2] | [20-30] | [°C) | Very High |
| | | | | | | [32.57-34.7] | [25-40] | [PSU] | |
| | | | | | | [0-54.67] | [0-70] | [m] | |
| 11 | *Pocillopora damicornis* | 312 | 7.18 | 97.81 | 26.50 | [7.06-16.97] | [-5-30] | [°C) | High |
| | | | | | | [34.45-34.99] | [25-40] | [PSU] | |
| | | | | | | [23.57-414.71] | [0-200] | [m] | |
| 13 | *Pomadasys panamensis* | 73 | 24.66 | 7.77 | 13.84 | [15.48-24.55] | [20-30] | [°C) | High |
| | | | | | | [33.73-34.97] | [25-40] | [PSU] | |
| | | | | | | [1.56-89.67] | [0-70] | [m] | |
| 14 | *Rhizoprionod on longurio* | 16 | 1.78 | 72.73 | 11.38 | [13.66-23.87] | [5-30] | [°C) | Very High |
| | | | | | | [32.8-34.98] | [30-40] | [PSU] | |
| | | | | | | [0-120] | [0-300] | [m] | |
| 15 | *Lepidochelys olivacea* | 6 | 33.33 | 6.38 | 14.58 | [6.52-15.13] | [5-30-] | [°C) | Very High |
| | | | | | | [34.6-34.99] | [15-40] | [PSU] | |
| | | | | | | [39.3-461.2] | [0-544] | [m] | |
| 16 | *Tripos fusus* | 5 | 9.09 | 17.86 | 12.74 | [6.48-16.1] | [-5-35] | [°C) | Very High |
| | | | | | | [34.58-34.99] | [0-40] | [PSU] | |
| | | | | | | [71.3-773] | [0-4000] | [m] | |

Validation of species-habitat environmental preferences demonstrated high biological coherence for all clusters meeting established criteria, evidencing the SOM model's capacity to capture ecologically relevant environmental gradients. Coastal shallow habitats (clusters 1, 5, 6, 9, 10, 13) exhibited the highest degree of validation, with six clusters characterized by species typical of warm tropical environments with low to moderate salinity. Species such as Pomadasys panamensis (clusters 1, 10, 13), Lutjanus guttatus (cluster 3), and Litopenaeus occidentalis (clusters 5, 6, 9) presented documented preference ranges in the literature that precisely coincided with environmental conditions observed in these habitats (Table 2). Particularly, the recurrence of P. panamensis and L. occidentalis in multiple coastal clusters possibly stems from spatial structuring of these shallow habitats that, although environmentally similar, could represent local variations in community composition or habitat use intensity.

Transition habitats (clusters 3, 4, 7, 8) showed solid correspondence with species of intermediate ecological amplitude, including Solenocera agassizii, Xiphopenaeus riveti, and Penaeus brevirostris. The presence of these species adequately reflects the ecotonal nature of these environments, which function as interface zones between shallow coastal conditions and deeper oceanic environments. Particularly relevant is the case of cluster 2, characterized by Mugil cephalus, a euryhaline species with tolerance to wide salinity ranges (10–40 PSU) and temperature (5–40°C), whose presence validates identification of this habitat as an estuarine-marine interface environment with high environmental variability.

Validation analysis also revealed differences in inference robustness according to cluster sample size. Clusters with higher number of occurrences ($n > 200$; clusters 1, 5, 8, 9) showed more consistent patterns and dominant species with higher composition percentages (21–45%), while clusters with fewer records ($n < 100$; clusters 2, 3, 7) exhibited lower relative dominance (14–15%) and greater associated uncertainty. This relationship between sampling effort and validation robustness underscores the need for stratified sampling designs that explicitly consider representativeness of all classified habitats.

## 4. Conclusions

The study demonstrated that integration of the ECOSAIC model and Self-Organizing Map (SOM) offers an effective solution for overcoming technical and logistical challenges in benthic habitat mapping in complex and poorly explored regions such as the Colombian Pacific. The results obtained align with recent advances in Deep Learning applied to marine studies that highlight the potential of hybrid approaches for improving both precision and ecological interpretability of benthic habitat classifications.

Personalized semantic compression and domain specialization enabled capturing key functional differences between biogeochemical and hydrogeomorphological variables, which translated into effective dimensional reduction and identification of 16 potential ecologically significant benthic habitat classes. The ECOSAIC architecture achieved substantial orthogonality between latent representations while preserving ecologically interpretable information, surpassing limitations of traditional dimensional reduction methods that frequently sacrifice biological meaning for mathematical efficiency.

The multi-level validation implemented, including MIC analysis of functional specialization, environmental correspondence with known oceanographic patterns, and biological validation through species occurrences, demonstrated the methodological robustness of the approach. The significant correlation between environmental gradients and species composition together with high community differentiation among habitats confirm that the classification captured ecologically relevant units beyond simple statistical groupings.

The high correspondence between environmental patterns and observed species distribution validates the model's predictive and functional capacity, supporting its utility for decision-making oriented toward marine management and conservation. The identified habitats reflect the characteristic dichotomy between coastal systems influenced by mangroves and rivers versus deep oceanic habitats, each with distinctive biogeochemical and hydrogeomorphological signatures that correspond with specialized biological communities documented in scientific literature.

The developed methodology is scalable and adaptable for application in distinct marine ecological contexts, particularly relevant for Eastern Tropical Pacific regions where the combination of high biodiversity, pronounced environmental gradients, and logistical limitations for sampling make development of efficient mapping tools critical. This work represents a significant advance in generating functional benthic maps that contribute to marine spatial planning and conservation under global environmental change scenarios, providing essential tools for management and conservation of marine species.

**Acknowledgements**

The authors are grateful to the Autoridad Nacional de Acuicultura y Pesca (AUNAP) and Instituto de Investigaciones Marinas y Costeras José Benito Vives de Andréis (INVEMAR) for providing valuable data.

## 5. References

Alfaro, A. C., & Alfaro, A. C. (2010). Effects of mangrove removal on benthic communities and sediment characteristics at Mangawhai Harbour, northern New Zealand. *ICES Journal of Marine Science*, *67*(6), 1087–1104. https://doi.org/10.1093/ICESJMS/FSQ034

Assis, J., Fernández Bejarano, S. J., Salazar, V. W., Schepers, L., Gouvêa, L., Fragkopoulou, E., Leclercq, F., Vanhoorne, B., Tyberghein, L., Serrão, E. A., Verbruggen, H., & De Clerck, O. (2024). Bio-ORACLE v3.0. Pushing marine data layers to the CMIP6 Earth System Models of climate change research. *Global Ecology and Biogeography*, *33*(4), e13813. https://doi.org/10.1111/GEB.13813

Baig, Y., Ma, H. R., Xu, H., & You, L. (2023). Autoencoder neural networks enable low dimensional structure analyses of microbial growth dynamics. *Nature Communications*, *14*(1), 1–17. https://doi.org/10.1038/S41467-023-43455-0

Baselga, A. (2010). Partitioning the turnover and nestedness components of beta diversity. *Global Ecology and Biogeography*, *19*(1), 134–143. https://doi.org/10.1111/J.1466-8238.2009.00490.X

Basher, Z., Bowden, D. A., & Costello, M. J. (2018). *GMED: Global Marine Environment Datasets for environment visualisation and species distribution modelling*. https://doi.org/10.5194/ESSD-2018-64

Benavides Martínez, I. F., Rueda, M., Ortíz Ferrin, O. O., Díaz-Ochoa, J. A., Castillo-Vargasmachuca, S., & Selvaraj, J. J. (2024). A novel approach for improving the spatiotemporal distribution modeling of marine benthic species by coupling a new GIS procedure with machine learning. *Deep Sea Research Part I: Oceanographic Research Papers*, *203*, 104222. https://doi.org/10.1016/J.DSR.2023.104222

Bolton, T., & Zanna, L. (2019). Applications of Deep Learning to Ocean Data Inference and Subgrid Parameterization. Journal of Advances in Modeling Earth Systems, 11(1), 376–399. https://doi.org/10.1029/2018MS001472

Brown, C. J., Smith, S. J., Lawton, P., & Anderson, J. T. (2011). Benthic habitat mapping: A review of progress towards improved understanding of the spatial ecology of the seafloor using acoustic techniques. *Estuarine, Coastal and Shelf Science*, *92*(3), 502–520. https://doi.org/10.1016/J.ECSS.2011.02.007

Chon, T. S., Jang, Y. H., Jung, N., Lee, K. E., Kwak, G. S., Kim, D. H., Sim, K. S., Lee, J. E., Min, J. H., & Park, Y. S. (2023). Spatial patterning of benthic macroinvertebrate communities using Geo-self-organizing map (Geo-SOM): A case study in the Nakdong River, South Korea. *Ecological Informatics*, *76*, 102148. https://doi.org/10.1016/J.ECOINF.2023.102148

Clarke, K. R., Gorley, R. N., Somerfield, P. J., & Warwick, R. M. (2014). *Change in marine communities: an approach to statistical analysis and interpretation*.

Costa, B. M., Battista, T. A., & Pittman, S. J. (2009). Comparative evaluation of airborne LiDAR and ship-based multibeam SoNAR bathymetry and intensity for mapping coral reef ecosystems. *Remote Sensing of Environment*, *113*(5), 1082–1100. https://doi.org/10.1016/J.RSE.2009.01.015

Covich, A. P., Austen, M. C., BÄRlocher, F., Chauvet, E., Cardinale, B. J., Biles, C. L., Inchausti, P., Dangles, O., Solan, M., Gessner, M. O., Statzner, B., & Moss, B. (2004). The Role of Biodiversity in the Functioning of Freshwater and Marine Benthic Ecosystems. *BioScience*, *54*(8), 767–775. https://doi.org/10.1641/0006-3568(2004)054[0767:TROBIT]2.0.CO;2

De Cáceres, M., Legendre, P., Wiser, S. K., & Brotons, L. (2012). Using species combinations in indicator value analyses. Methods in Ecology and Evolution, 3(6), 973–982. https://doi.org/10.1111/j.2041-210X.2012.00246.x

Devictor, V., Julliard, R., & Jiguet, F. (2008). Distribution of specialist and generalist species along spatial gradients of habitat disturbance and fragmentation. *Oikos*, *117*(4), 507–514. https://doi.org/10.1111/J.0030-1299.2008.16215.X

Díaz-Merlano, J. M. (2007). Deltas and Estuaries of Colombia [*Deltas y Estuarios de Colombia]*. https://comunidadplanetaazul.com/ecolibros/deltas-y-estuarios-de-colombia/

Duarte, C. M., Gattuso, J. P., Hancke, K., Gundersen, H., Filbee-Dexter, K., Pedersen, M. F., Middelburg, J. J., Burrows, M. T., Krumhansl, K. A., Wernberg, T., Moore, P., Pessarrodona, A., Ørberg, S. B., Pinto, I. S., Assis, J., Queirós, A. M., Smale, D. A., Bekkby, T., Serrão, E. A., & Krause-Jensen, D. (2022). Global estimates of the extent and production of macroalgal forests. *Global Ecology and Biogeography*, *31*(7), 1422–1439. https://doi.org/10.1111/GEB.13515

Dufrêne, M., & Legendre, P. (1997). Species assemblages and indicator species: The need for a flexible asymmetrical approach. *Ecological Monographs*, *67*(3), 345. **https://doi.org/10.1890/0012-9615(1997)067[0345:SAAIST]2.0.CO;2**

Evans, S. N., Konzewitsch, N., Hovey, R. K., Kendrick, G. A., & Bellchambers, L. M. (2025). Selecting the best habitat mapping technique: a comparative assessment for fisheries management in Exmouth Gulf. *Frontiers in Marine Science*, *12*, 1570277. https://doi.org/10.3389/FMARS.2025.1570277/BIBTEX

Falasca, F., Perezhogin, P., & Zanna, L. (2024). Data-driven dimensionality reduction and causal inference for spatiotemporal climate fields. *Physical Review E*, *109*(4), 044202. https://doi.org/10.1103/PhysRevE.109.044202

Galparsoro, I., Borja, A., & Uyarra, M. C. (2014). Mapping ecosystem services provided by benthic habitats in the European North Atlantic Ocean. *Frontiers in Marine Science*, *1*(JUL), 96488. https://doi.org/10.3389/FMARS.2014.00023/BIBTEX

Game, C. A., Thompson, M. B., & Finlayson, G. D. (2024). Machine learning for non-experts: A more accessible and simpler approach to automatic benthic habitat classification. *Ecological Informatics*, *81*, 102619. https://doi.org/10.1016/J.ECOINF.2024.102619

GBIF.org (03 November 2025) GBIF Occurrence Download https://doi.org/10.15468/dl.rt2v73

Gómez, E., & Bernal, G. (2013). Influence of the environmental characteristics of mangroves on recent benthic foraminifera in the Gulf of Urabá, Colombian Caribbean. Marine Sciences. [Influencia de las características ambientales de los manglares sobre foraminíferos bénticos recientes en el golfo de Urabá, Caribe colombiano. *Ciencias Marinas*], *39*(1), 69–82. https://doi.org/10.7773/CM.V39I1.2175

González-Valdivia, N., Ochoa-Gaona, S., Pozo, C., Gordon Ferguson, B., Rangel-Ruiz, L. J., Arriaga-Weiss, S. L., Ponce-Mendoza, A., & Kampichler, C. (2011). Ecological indicators of habitat and biodiversity in a neotropical landscape: a multitaxonomic perspective. Journal of Tropical Biology [Indicadores ecológicos de hábitat y biodiversidad en un paisaje neotropical: perspectiva multitaxonómica. Revista de Biología Tropical], 59(3), 1433–1451. http://www.scielo.sa.cr/scielo.php?script=sci_arttext&pid=S0034-77442011000300039&lng=en&nrm=iso&tlng=es

Guisan, A., & Thuiller, W. (2005). Predicting species distribution: Offering more than simple habitat models. *Ecology Letters*, *8*(9), 993–1009. https://doi.org/10.1111/J.1461-0248.2005.00792.X

Hasan, R. C., Ierodiaconou, D., & Monk, J. (2012). Evaluation of Four Supervised Learning Methods for Benthic Habitat Mapping Using Backscatter from Multi-Beam Sonar. Remote Sensing 2012, Vol. 4, Pages 3427-3443, 4(11), 3427–3443. https://doi.org/10.3390/RS4113427

Ham, Y. G., Joo, Y. S., Kim, J. H., & Lee, J. G. (2024). Partial-convolution-implemented generative adversarial network for global oceanic data assimilation. Nature Machine Intelligence 2024 6:7, 6(7), 834–843. https://doi.org/10.1038/s42256-024-00867-x

Heimbach, P., O'Donncha, F., Smith, T., Garcia-Valdecasas, J. M., Arnaud, A., & Wan, L. (2025). *Crafting the Future: Machine learning for ocean forecasting*. https://doi.org/10.5194/SP-5-OPSR-22-2025

Hersbach, H., Bell, B., Berrisford, P., Biavati, G., Horányi, A., Muñoz-Sabater, J., Nicolas, J., Peubey, C., Radu, R., Rozum, I., Schepers, D., Simmons, A., Soci, C., Dee, D., & Thépaut, J.-N. (2023). *ERA5 hourly data on single levels from 1940 to present*. https://doi.org/10.24381/cds.adbb2d47

Hijmans, R. J. (2023). Geographic Data Analysis and Modeling [R package raster version 3.6-23]. https://cran.r-project.org/web/packages/raster/raster.pdf

Hilmi, N., Sutherland, M., Farahmand, S., Haraldsson, G., van Doorn, E., Ernst, E., Wisz, M. S., Claudel Rusin, A., Elsler, L. G., & Levin, L. A. (2023). Deep sea nature-based solutions to climate change. *Frontiers in Climate*, *5*, 1169665. https://doi.org/10.3389/FCLIM.2023.1169665/XML

Hodson, T. O. (2022). Root-mean-square error (RMSE) or mean absolute error (MAE): when to use them or not. *Geoscientific Model Development*, *15*(14), 5481–5487. https://doi.org/10.5194/GMD-15-5481-2022

Horé, A., & Ziou, D. (2010). Image quality metrics: PSNR vs. SSIM. *Proceedings - International Conference on Pattern Recognition*, 2366–2369. https://doi.org/10.1109/ICPR.2010.579

Instituto Geográfico Agustín Codazzi - IGAC. (2016). Vector cartography at a scale of 1:500,000 with total coverage of the Republic of Colombia [Cartografía vectorial a escala 1:500.000 con cobertura total de la República de Colombia]. https://www.colombiaenmapas.gov.co/?e=-83.15195911133215,0.4332944028567231,-70.5176817675855,9.302075086892238,4686&b=igac&u=0&t=23&servicio=204

Instituto de Investigaciones Marinas y Costeras - INVEMAR. (2023). SIAM Biological Records [SIAM Registros Biológicos]. https://acceso-datos-ambientales-invemar.hub.arcgis.com/maps/INVEMAR::siam-registrosbiologicos-1/about

Julliard, R., Clavel, J., Devictor, V., Jiguet, F., & Couvet, D. (2006). Spatial segregation of specialists and generalists in bird communities. *Ecology Letters*, *9*(11), 1237–1244. https://doi.org/10.1111/J.1461-0248.2006.00977.X

Lecours, V., Brown, C. J., Devillers, R., Lucieer, V. L., & Edinger, E. N. (2016). Comparing Selections of Environmental Variables for Ecological Studies: A Focus on Terrain Attributes. *PLOS ONE*, *11*(12), e0167128. https://doi.org/10.1371/JOURNAL.PONE.0167128

Lecours, V., Devillers, R., Simms, A. E., Lucieer, V. L., & Brown, C. J. (2017). Towards a framework for terrain attribute selection in environmental studies. *Environmental Modelling & Software*, *89*, 19–30. https://doi.org/10.1016/J.ENVSOFT.2016.11.027

Legendre, P., & Legendre, L. (2012). Numerical Ecology. *Developments in Environmental Modelling*, *24*, 337–424. http://www.sciencedirect.com/science/article/pii/B9780444538680500083

Li, G., & Cao, G. (2025). Generative adversarial models for extreme geospatial downscaling. *International Journal of Applied Earth Observation and Geoinformation*, *139*. https://doi.org/10.1016/j.jag.2025.104541

Liu, Y., Weisberg, R. H., & Mooers, C. N. K. (2006). Performance evaluation of the self-organizing map for feature extraction. Journal of Geophysical Research: Oceans, 111(5). **https://doi.org/10.1029/2005JC003117**

Marburg, A., & Bigham, K. (2016). Deep learning for benthic fauna identification. *OCEANS 2016 MTS/IEEE Monterey, OCE 2016*. https://doi.org/10.1109/OCEANS.2016.7761146

Misiuk, B., Brown, C. J., Robert, K., & Lacharité, M. (2020). Harmonizing Multi-Source Sonar Backscatter Datasets for Seabed Mapping Using Bulk Shift Approaches. *Remote Sensing 2020, Vol. 12, Page 601*, *12*(4), 601. https://doi.org/10.3390/RS12040601

Mohamed, H., Nadaoka, K., & Nakamura, T. (2020). Towards Benthic Habitat 3D Mapping Using Machine Learning Algorithms and Structures from Motion Photogrammetry. *Remote Sensing 2020, Vol. 12, Page 127*, *12*(1), 127. https://doi.org/10.3390/RS12010127

Naeem, S., Prager, C., Weeks, B., Varga, A., Flynn, D. F. B., Griffin, K., Muscarella, R., Palmer, M., Wood, S., & Schuster, W. (2016). Biodiversity as a multidimensional construct: a review, framework and case study of herbivory's impact on plant biodiversity. *Proceedings of the Royal Society B: Biological Sciences*, *283*(1844), 20153005. https://doi.org/10.1098/RSPB.2015.3005

National Research Council. (2009). Science at Sea: Meeting Future Oceanographic Goals with a Robust Academic Research Fleet. *Science at Sea: Meeting Future Oceanographic Goals with a Robust Academic Research Fleet*, 1–120. https://doi.org/10.17226/12775

Navarro, C., Janc, A., Lassalle, G., Lambert, P., & Dambrine, C. (2023). From the modeling of diadromous species' marine distributions to the characterization of their current and future marine habitats. *Frontiers in Marine Science*, *10*, 1241969. https://doi.org/10.3389/FMARS.2023.1241969/BIBTEX

Oviedo-Barrero, F., Niño-Pinzón, D. C., Aguirre-Tapiero, M. del P., Pantoja-López, D. N., & Sánchez-Manco, L. (2020). *Capítulo I – Particularidades geográficas de la Cuenca Pacífica Colombiana. En Compilación Oceanográfica de la Cuenca Pacífica Colombiana*. https://www.researchgate.net/publication/353417938_Capitulo_I_-_Particularidades_geograficas_de_la_Cuenca_Pacifica_Colombiana_En_Compilacion_Oceanografica_de_la_Cuenca_Pacifica_Colombiana_II_Pp_34-64_Direccion_General_Maritima_Bogota_D_C_Editorial_DIM

RAP Pacífico. (2022). Pacific Regional Strategic Plan [*Plan Estratégico Regional Pacífico]*. https://rap-pacifico.gov.co/wp-content/uploads/2022/12/PER-PACIFICO.pdf

Restrepo, J. D., & Kjerfve, B. (2004). The Pacific and Caribbean Rivers of Colombia: Water Discharge, Sediment Transport and Dissolved Loads. *Environmental Geochemistry in Tropical and Subtropical Environments*, 169–187. https://doi.org/10.1007/978-3-662-07060-4_14

Restrepo, J. D., Kjerfve, B., Correa, I. D., & González, J. (2002). Morphodynamics of a high discharge tropical delta, San Juan River, Pacific coast of Colombia. *Marine Geology*, *192*(4), 355–381. https://doi.org/10.1016/S0025-3227(02)00579-0

Rosales-Estrella, D. L., Portilla-Cabrera, C. V., Guzmán-Alvis, Á. I., Enterline, C., Gil, A. D., Rueda, M., & Benavides-Martínez, I. F. (2025). Machine learning based mapping of physicochemical attributes in the Colombian Pacific seafloor. *Earth Science Informatics*, *18*(3), 1–20. https://doi.org/10.1007/S12145-025-01905-X/FIGURES/7

Salgado, A., He, Y., Radke, J., Ganguly, A. R., & Gonzalez, M. C. (2024). Dimension reduction approach for understanding resource-flow resilience to climate change. *Communications Physics*, *7*(1), 1–12. https://doi.org/10.1038/S42005-024-01664-Z;SUBJMETA

Song, M. Y., Hwang, H. J., Kwak, I. S., Ji, C. W., Oh, Y. N., Youn, B. J., & Chon, T. S. (2007). Self-organizing mapping of benthic macroinvertebrate communities implemented to community assessment and water quality evaluation. *Ecological Modelling*, *203*(1–2), 18–25. https://doi.org/10.1016/J.ECOLMODEL.2006.04.027

Sørensen, O. J. R., van Rijn, I., Einbinder, S., Nativ, H., Scheinin, A., Zemah-Shamir, Z., Bigal, E., Livne, L., Tsemel, A., Bialik, O. M., Papeer, G., Tchernov, D., & Makovsky, Y. (2025). Bridging the gap in deep seafloor management: Ultra fine-scale ecological habitat characterization of large seascapes. *Remote Sensing in Ecology and Conservation*, *11*(4), 472–489. https://doi.org/10.1002/RSE2.70002;WEBSITE:WEBSITE:ZSLPUBLICATIONS;REQUESTEDJOURNAL:JOURNAL:20563485;WGROUP:STRING:PUBLICATION

Selvaraj, J. J., & Gallego Pérez, B. E. (2023). An enhanced approach to mangrove forest analysis in the Colombian Pacific coast using optical and SAR data in Google Earth Engine. Remote Sensing

Applications: Society and Environment, 30, 100938. https://doi.org/10.1016/J.RSASE.2023.100938

Spring, D. L., & Williams, G. J. (2023). Influence of upwelling on coral reef benthic communities: a systematic review and meta-analysis. Proceedings of the Royal Society B: Biological Sciences, 290(1995), 20230023. https://doi.org/10.1098/RSPB.2023.0023

Sun, K., Cui, W., & Chen, C. (2021). Review of Underwater Sensing Technologies and Applications. *Sensors 2021, Vol. 21, Page 7849*, *21*(23), 7849. https://doi.org/10.3390/S21237849

Tsikopoulou, I., Nasi, F., & Bremner, J. (2024). Editorial: The importance of understanding benthic ecosystem functioning. *Frontiers in Marine Science*, *11*, 1470915. https://doi.org/10.3389/FMARS.2024.1470915/BIBTEX

Üstek, İ., Arana-Catania, M., Farr, A., & Petrunin, I. (2024). Deep Autoencoders for Unsupervised Anomaly Detection in Wildfire Prediction. *Earth and Space Science*, *11*(11), e2024EA003997. https://doi.org/10.1029/2024EA003997

Wang, Z., Guo, J., Zhang, S., & Xu, N. (2025). Marine object detection in forward-looking sonar images via semantic-spatial feature enhancement. *Frontiers in Marine Science*, *12*, 1539210. https://doi.org/10.3389/FMARS.2025.1539210/BIBTEX

Wicaksono, P., Aryaguna, P. A., & Lazuardi, W. (2019). Benthic Habitat Mapping Model and Cross Validation Using Machine-Learning Classification Algorithms. *Remote Sensing 2019, Vol. 11, Page 1279*, *11*(11), 1279. https://doi.org/10.3390/RS11111279

Xu, P., Ma, Y., Lu, W., Li, M., Zhao, W., & Dai, Z. (2025). Multi-objective optimization in machine learning assisted materials design and discovery. *J. Mater. Inf. 2025, 5, 26.* , *5*(2), N/A-N/A. https://doi.org/10.20517/JMI.2024.108

Yang, L., Lu, W., He, L., Huang, H., Liu, Z., Zhang, Y., Ma, Y., Yu, J., Zuo, G., Ai, Y., Liu, C., & Xu, Y. (2025a). Mapping benthic habitats in Bohai Bay, China. *Scientific Reports*, *15*(1), 1–11. https://doi.org/10.1038/S41598-025-02091-Y;SUBJMETA=158,631,670,672,704,826,829;KWRD=BIODIVERSITY,CONSERVATION+BIOLOGY,MARINE+BIOLOGY,OCEAN+SCIENCES

Yang, L., Lu, W., He, L., Huang, H., Liu, Z., Zhang, Y., Ma, Y., Yu, J., Zuo, G., Ai, Y., Liu, C., & Xu, Y. (2025b). Mapping benthic habitats in Bohai Bay, China. *Scientific Reports*, *15*(1), 1–11. https://doi.org/10.1038/S41598-025-02091-Y;SUBJMETA

Zhao, S., Cui, J., Sheng, Y., Dong, Y., Liang, X., Chang, E. I. C., & Xu, Y. (2021). Large Scale Image Completion via Co-Modulated Generative Adversarial Networks. *ICLR 2021 - 9th International Conference on Learning Representations*. https://arxiv.org/pdf/2103.10428

**Supplementary Material**

Table S1. Performance comparison of spatial data gap-filling methods for environmental variables using longitudinal cross-validation (n = 3 folds). Values presented as mean ± standard deviation. L1 = Mean Absolute Error (lower values indicate better performance); L2 = Mean Squared Error (lower values indicate better performance); PSNR = Peak Signal-to-Noise Ratio (higher values indicate better performance); SSIM = Structural Similarity Index (higher values indicate better performance, range 0–1).

| Variable | Method | L1 (MAE) | L2 (MSE) | PSNR (dB) | SSIM |
|---|---|---|---|---|---|
| Phosphate [mol.m$^{-3}$] | C-GAN | 0.0527 ± 0.0034 | 0.0049 ± 0.0007 | 23.18 ± 0.62 | 0.928 ± 0.026 |
| | U-Net | 0.0514 ± 0.0011 | 0.0052 ± 0.0024 | 23.32 ± 2.07 | 0.928 ± 0.023 |
| | Kriging | 0.2182 ± 0.039 | 0.0701 ± 0.024 | 11.85 ± 1.70 | 0.265 ± 0.218 |
| Iron [umol.m$^{-3}$] | C-GAN | 0.0523 ± 0.0073 | 0.0056 ± 0.0012 | 22.61 ± 0.86 | 0.731 ± 0.044 |
| | U-Net | 0.1008 ± 0.0536 | 0.0197 ± 0.018 | 18.87 ± 3.94 | 0.477 ± 0.291 |
| | Kriging | 0.1607 ± 0.022 | 0.0396 ± 0.007 | 14.09 ± 0.75 | 0.231 ± 0.020 |
| Bottom light | C-GAN | 0.0591 ± 0.026 | 0.0161 ± 0.010 | 18.75 ± 2.65 | 0.684 ± 0.073 |
| | U-Net | 0.1289 ± 0.070 | 0.0337 ± 0.024 | 15.79 ± 2.94 | 0.514 ± 0.228 |
| | Kriging | 0.1538 ± 0.030 | 0.0420 ± 0.009 | 13.87 ± 0.99 | 0.085 ± 0.035 |
| Nitrate [mol.m$^{-3}$] | C-GAN | 0.0518 ± 0.0066 | 0.0050 ± 0.0012 | 23.11 ± 0.98 | 0.915 ± 0.046 |
| | U-Net | 0.0799 ± 0.024 | 0.0130 ± 0.007 | 19.77 ± 3.13 | 0.843 ± 0.073 |
| | Kriging | 0.2286 ± 0.039 | 0.0751 ± 0.025 | 11.53 ± 1.65 | 0.260 ± 0.219 |
| Dissolved oxygen [mmol.m$^{-3}$] | C-GAN | 0.0716 ± 0.0073 | 0.0102 ± 0.0026 | 20.06 ± 1.13 | 0.894 ± 0.025 |
| | U-Net | 0.0725 ± 0.009 | 0.0108 ± 0.003 | 19.85 ± 1.26 | 0.861 ± 0.070 |
| | Kriging | 0.2623 ± 0.041 | 0.0926 ± 0.025 | 10.53 ± 1.36 | 0.249 ± 0.224 |
| pH | C-GAN | 0.0511 ± 0.0095 | 0.0042 ± 0.0016 | 24.11 ± 1.56 | 0.526 ± 0.136 |
| | U-Net | 0.1019 ± 0.062 | 0.0270 ± 0.025 | 17.95 ± 4.57 | 0.611 ± 0.128 |
| | Kriging | 0.1956 ± 0.137 | 0.0831 ± 0.062 | 20.55 ± 16.19 | 0.572 ± 0.298 |
| Salinity [PSU] | C-GAN | 0.0598 ± 0.015 | 0.0086 ± 0.004 | 21.08 ± 2.00 | 0.848 ± 0.042 |
| | U-Net | 0.1112 ± 0.033 | 0.0248 ± 0.011 | 16.65 ± 2.44 | 0.725 ± 0.086 |
| | Kriging | 0.1942 ± 0.010 | 0.0493 ± 0.009 | 13.15 ± 0.83 | 0.320 ± 0.169 |
| Silicate [mol.m− 3] | C-GAN | 0.0396 ± 0.008 | 0.0041 ± 0.0014 | 24.16 ± 1.44 | 0.880 ± 0.032 |
| | U-Net | 0.1091 ± 0.051 | 0.0217 ± 0.015 | 17.53 ± 2.69 | 0.601 ± 0.039 |
| | Kriging | 0.1459 ± 0.051 | 0.0382 ± 0.016 | 14.68 ± 2.28 | 0.305 ± 0.176 |
| Temperature [°C] | C-GAN | 0.0543 ± 0.006 | 0.0060 ± 0.002 | 22.41 ± 1.33 | 0.927 ± 0.031 |
| | U-Net | 0.0423 ± 0.008 | 0.0034 ± 0.001 | 24.79 ± 1.09 | 0.930 ± 0.013 |
| | Kriging | 0.2234 ± 0.031 | 0.0694 ± 0.019 | 11.77 ± 1.37 | 0.264 ± 0.230 |
| Photosynthetically active radiation [Einstein/m²/day] | C-GAN | 0.0540 ± 0.0027 | 0.0049 ± 0.0004 | 23.08 ± 0.37 | 0.757 ± 0.051 |
| | U-Net | 0.1040 ± 0.037 | 0.0194 ± 0.013 | 18.30 ± 3.40 | 0.577 ± 0.100 |
| | Kriging | 0.1092 ± 0.025 | 0.0214 ± 0.012 | 17.46 ± 2.65 | 0.631 ± 0.166 |
| Organic matter [%] | C-GAN | 0.0895 ± 0.0067 | 0.0142 ± 0.0025 | 18.53 ± 0.74 | 0.597 ± 0.038 |
| | U-Net | 0.1660 ± 0.081 | 0.0473 ± 0.041 | 14.82 ± 3.58 | 0.250 ± 0.196 |
| | Kriging | 0.1208 ± 0.015 | 0.0215 ± 0.005 | 16.80 ± 1.03 | 0.261 ± 0.081 |
| Chlorophyll-a [mg·m⁻³] | C-GAN | 0.0517 ± 0.0046 | 0.0051 ± 0.0005 | 22.91 ± 0.45 | 0.877 ± 0.031 |
| | U-Net | 0.0497 ± 0.014 | 0.0055 ± 0.003 | 23.33 ± 2.49 | 0.843 ± 0.058 |
| | Kriging | 0.1913 ± 0.039 | 0.0496 ± 0.017 | 13.33 ± 1.62 | 0.266 ± 0.193 |
| Phytoplankton [umol.m⁻³] | C-GAN | 0.0490 ± 0.0074 | 0.0044 ± 0.0011 | 23.64 ± 1.02 | 0.894 ± 0.029 |
| | U-Net | 0.0538 ± 0.014 | 0.0055 ± 0.002 | 23.01 ± 2.07 | 0.841 ± 0.097 |
| | Kriging | 0.1890 ± 0.036 | 0.0500 ± 0.015 | 13.20 ± 1.28 | 0.285 ± 0.171 |
| Bottom utilized oxygen [ml/l] | C-GAN | 0.0414 ± 0.018 | 0.0054 ± 0.005 | 24.76 ± 4.09 | 0.823 ± 0.100 |
| | U-Net | 0.1041 ± 0.049 | 0.0203 ± 0.015 | 18.46 ± 3.84 | 0.660 ± 0.137 |
| | Kriging | 0.1725 ± 0.013 | 0.0605 ± 0.020 | 12.41 ± 1.43 | 0.384 ± 0.084 |
| Primary productivity [g·m⁻³·day⁻¹] | C-GAN | 0.0501 ± 0.015 | 0.0064 ± 0.003 | 22.36 ± 1.95 | 0.839 ± 0.027 |
| | U-Net | 0.0717 ± 0.042 | 0.0120 ± 0.010 | 20.96 ± 4.02 | 0.785 ± 0.106 |
| | Kriging | 0.1982 ± 0.018 | 0.0570 ± 0.013 | 12.55 ± 0.94 | 0.179 ± 0.138 |
| Benthic current velocity [m·s⁻¹] | C-GAN | 0.0525 ± 0.012 | 0.0048 ± 0.002 | 23.67 ± 2.17 | 0.732 ± 0.069 |
| | U-Net | 0.0903 ± 0.059 | 0.0167 ± 0.018 | 20.63 ± 5.01 | 0.540 ± 0.297 |
| | Kriging | 0.1619 ± 0.075 | 0.0444 ± 0.033 | 14.78 ± 3.30 | 0.181 ± 0.076 |
| Runoff [m] | C-GAN | 0.0300 ± 0.005 | 0.0030 ± 0.002 | 26.10 ± 3.12 | 0.821 ± 0.035 |
| | U-Net | 0.1337 ± 0.037 | 0.0306 ± 0.013 | 15.52 ± 1.81 | 0.282 ± 0.221 |
| | Kriging | 0.1694 ± 0.062 | 0.0464 ± 0.027 | 14.66 ± 3.92 | 0.264 ± 0.211 |
| 10 m v-component of wind [m.s− 1] | C-GAN | 0.0373 ± 0.008 | 0.0030 ± 0.001 | 25.59 ± 1.68 | 0.855 ± 0.067 |
| | U-Net | 0.0897 ± 0.031 | 0.0124 ± 0.007 | 19.65 ± 2.16 | 0.698 ± 0.051 |
| | Kriging | 0.1568 ± 0.055 | 0.0390 ± 0.021 | 15.20 ± 3.54 | 0.544 ± 0.169 |
| Substrate hardness [%] | C-GAN | 0.0936 ± 0.012 | 0.0141 ± 0.004 | 18.69 ± 1.23 | 0.551 ± 0.060 |
| | U-Net | 0.1640 ± 0.026 | 0.0393 ± 0.012 | 14.25 ± 1.26 | 0.283 ± 0.113 |
| | Kriging | 0.1476 ± 0.028 | 0.0327 ± 0.011 | 15.15 ± 1.66 | 0.167 ± 0.115 |
| Particle size [phi] | C-GAN | 0.0557 ± 0.011 | 0.0077 ± 0.004 | 21.83 ± 2.64 | 0.717 ± 0.047 |
| | U-Net | 0.1266 ± 0.054 | 0.0346 ± 0.017 | 15.29 ± 2.61 | 0.514 ± 0.164 |
| | Kriging | 0.1119 ± 0.040 | 0.0215 ± 0.011 | 17.05 ± 3.39 | 0.402 ± 0.175 |
| Euphotic layer depth [m] | C-GAN | 0.0366 ± 0.004 | 0.0025 ± 0.0005 | 26.18 ± 0.92 | 0.862 ± 0.036 |
| | U-Net | 0.0577 ± 0.015 | 0.0074 ± 0.004 | 22.11 ± 2.73 | 0.817 ± 0.096 |
| | Kriging | 0.1507 ± 0.049 | 0.0327 ± 0.018 | 15.82 ± 3.19 | 0.489 ± 0.289 |
| Light attenuation coefficient [m− 1] | C-GAN | 0.0380 ± 0.007 | 0.0030 ± 0.001 | 25.64 ± 1.84 | 0.842 ± 0.074 |
| | U-Net | 0.1048 ± 0.023 | 0.0210 ± 0.008 | 17.16 ± 1.90 | 0.512 ± 0.090 |
| | Kriging | 0.1683 ± 0.053 | 0.0406 ± 0.019 | 14.33 ± 1.86 | 0.365 ± 0.253 |

Table S2. Environmental preference ranges of indicator species

| **Specie** | **Temperature [°C]** | **Source** | **Salinity [PSU]** | **Sourse** | **Depth [m]** | **Sourse** |
|---|---|---|---|---|---|---|
| *Pocilloporadami cornis* | [-5-30] | https://obis.org/taxon/206953 | [25-40] | https://obis.org/taxon/206953 | [0-200] | https://waterworlds.info/marine-species/cauliflower-coral-pocillopora-damicornis/ |
| *Penaeus brevirostris* | [15-27] | https://www.sealifebase.se/country/CountrySpeciesSummary.php?id=14597&c_code=218 | [25-35] | https://obis.org/taxon/584943 | [10-450] | https://www.nature.com/articles/s41598-024-71029-7 |
| *Lepidochelys olivacea* | [5-30] | https://obis.org/taxon/220293 | [15-40] | https://obis.org/taxon/220293 | [0-544] | https://www.int-res.com/articles/meps2006/323/m |
| | | | | | | 323p253.pdf |
| *Pomadasyspanamensis* | [20-30] | https://obis.org/taxon/273509 | [25-40] | https://obis.org/taxon/273509 | [0-70] | https://obis.org/taxon/273509 |
| *Litopenaeus occidentalis* | [21.3-28.6] | https://www.sealifebase.se/summary/Penaeus-occidentalis.html | [25-35] | https://obis.org/taxon/377456 | [2-160] | https://www.sealifebase.se/summary/Penaeus-occidentalis.html |
| *Mugil cephalus* | [5-30] | https://obis.org/taxon/126983 | [10-40] | https://obis.org/taxon/126983 | [0-300] | https://obis.org/taxon/126983 |
| *Lutjanus guttatus* | [15-30] | https://obis.org/taxon/276539 | [25-40] | https://obis.org/taxon/276539 | [0-170] | https://obis.org/taxon/276539 |
| *Solenocera agassizii* | [20-30] | https://obis.org/taxon/377658 | [25-35] | https://obis.org/taxon/377658 | [20-500] | https://obis.org/taxon/377658 |
| *Rhizoprionodon longurio* | [5-30] | https://obis.org/taxon/271327 | [30-40] | https://obis.org/taxon/271327 | [0-300] | https://obis.org/taxon/271327 |
| *Tripos fusus* | [-5-35] | https://obis.org/taxon/840626 | [0-40] | https://obis.org/taxon/840626 | [0-4000] | https://obis.org/taxon/840626 |
| *Xiphopenaeus riveti* | [10-30] | https://www.sciencedirect.com/science/article/pii/S2352485525000507#fig0010 | [25-35] | https://obis.org/taxon/585446 | [2-80] | https://www.nature.com/articles/s41598-024-71029-7 |